\documentclass[12pt]{amsart}

\input{diagrams.tex}
\usepackage{amsmath, amsthm, amssymb}

 \theoremstyle{plain}
\newtheorem{thm}{Theorem}[section]
\newtheorem{cor}[thm]{Corollary}
\newtheorem{lem}[thm]{Lemma}
\newtheorem{prop}[thm]{Proposition}

\theoremstyle{definition}
\newtheorem{defi}[thm]{Definition}
\newtheorem{conj}[thm]{Conjecture}
\newtheorem{nota}[thm]{Notation}
\newtheorem{rem}[thm]{Remark}
\newtheorem{rems}[thm]{Remarks}
\newtheorem{exa}[thm]{Example}
\newtheorem{exas}[thm]{Examples}
\newtheorem{sit}[]{}

\newcommand{\brem}{\begin{rem}}
\newcommand{\brems}{\begin{rems}}
\newcommand{\erem}{\end{rem}}
\newcommand{\erems}{\end{rems}}
\newcommand{\bexa}{\begin{exa}}
\newcommand{\bexas}{\begin{exas}}
\newcommand{\eexa}{\end{exa}}
\newcommand{\eexas}{\end{exas}}
\newcommand{\bdefi}{\begin{defi}}
\newcommand{\edefi}{\end{defi}}
\newcommand{\bdefis}{\begin{defis}}
\newcommand{\edefis}{\end{defis}}
\newcommand{\bcor}{\begin{cor}}
\newcommand{\ecor}{\end{cor}}
\newcommand{\blem}{\begin{lem}}
\newcommand{\elem}{\end{lem}}
\newcommand{\bconv}{\begin{conv}}
\newcommand{\econv}{\end{conv}}
\newcommand{\bconj}{\begin{conj}}
\newcommand{\econj}{\end{conj}}
\newcommand{\bprop}{\begin{prop}}
\newcommand{\eprop}{\end{prop}}
\newcommand{\bthm}{\begin{thm}}
\newcommand{\ethm}{\end{thm}}
\newcommand{\bnota}{\begin{nota}}
\newcommand{\enota}{\end{nota}}
\newcommand{\bsit}{\begin{sit}}
\newcommand{\esit}{\end{sit}}
\newcommand{\be}{\begin{eqnarray}}
\newcommand{\ee}{\end{eqnarray}}
\newcommand{\bproof}{\begin{proof}}
\newcommand{\eproof}{\end{proof}}

\def\ba{\begin{array}}
\def\ea{\end{array}}

\def\cF{{\mathcal F}}
\def\cH{{\mathcal H}}
\def\cL{{\mathcal L}}

\def\cD{{\mathcal D}}
\def\cU{{\mathcal U}}
\def\cS{{\mathcal S}}

\def\cP{{\mathcal P}}
\def\cE{{\mathcal E}}
\def\cV{{\mathcal V}}

\newcommand{\card}{\operatorname{card}}

\newcommand{\spec}{\operatorname{spec}}

\newcommand{\Symb}{\operatorname{\sigma}}
\newcommand{\Res}{\operatorname{res}}

\newcommand{\Char}{\operatorname{Char}}
\newcommand{\cchar}{\operatorname{char}}

\newcommand{\ind}{\operatorname{ind}}

\newcommand{\Aut}{{\operatorname{Aut}}}

\newcommand{\rk}{{\operatorname{rk}}}
\newcommand{\spann}{{\operatorname{span}}}

\newcommand{\res}{{\operatorname{res}}}

\newcommand{\Tr}{{\operatorname{Tr}}}

\newcommand{\GF}{{\operatorname{GF}}}

\newcommand{\Conv}{{\operatorname{Conv}}}
\newcommand{\lcm}{{\operatorname{lcm}}}
\newcommand{\SL}{{\bf {SL}}}
\newcommand{\GL}{{\bf {GL}}}

\newcommand{\ord}{{\operatorname{ord}}}
\newcommand{\mord}{{\operatorname{multi-ord}}}
\newcommand{\mult}{{\operatorname{mult}}}

\newcommand{\End}{{\operatorname{End}}}
\newcommand{\spa}{{\operatorname{span}}}
\newcommand{\supp}{{\operatorname{supp}}}
\newcommand{\Ann}{{\operatorname{Ann}}}
\newcommand{\charpol}{{\operatorname{CharPoly}}}

\newcommand{\A}{{\mathbb A}}

\newcommand{\Z}{{\mathbb Z}}
\newcommand{\N}{{\mathbb N}}
\newcommand{\NO}{{\mathbb N}_{\rm odd}}

\newcommand{\NC}{{\mathbb N}_{{\rm co}(p)}}

\newcommand{\T}{{\mathbb T}}
\newcommand{\F}{{\mathbb F}}
\newcommand{\G}{{\Gamma}}
\newcommand{\D}{{\Delta}}

\begin{document}

\title[Convolution equations on lattices and their  periodic solutions]
{Convolution equations on lattices:\\  periodic solutions with
values in \\a prime characteristic field}

\author{Mikhail Zaidenberg}
\address{Universit\'e
Grenoble I, Institut Fourier, UMR 5582 CNRS-UJF, BP 74, 38402 St.\
Martin d'H\`eres c\'edex, France} \email{zaidenbe@ujf-grenoble.fr}

\thanks{
{\bf Acknowledgements:} This work  partially was done during the
author's visit to the MPIM at Bonn. The author thanks this
institution for a generous support and excellent working
conditions.}

\thanks{
\mbox{\hspace{11pt}}{\it 1991 Mathematics Subject
Classification}: 11B39, 11T06, 11T99, 31C05, 37B15, 43A99.\\
\mbox{\hspace{11pt}}{\it Key words}: cellular automaton,
Chebyshev-Dickson polynomial, convolution operator, lattice,
sublattice, finite field, discrete Fourier transform, discrete
harmonic function, pluri-periodic function.}

\date{}


\begin{abstract}
These notes are inspired by the theory of cellular automata. The
latter aims, in particular, to provide a model for inter-cellular
or inter-molecular interactions. A linear cellular automaton on a
lattice $\Lambda$ is a discrete dynamical system generated by a
convolution operator $\D_a:f\longmapsto f*a$ with kernel $a$
concentrated in the nearest neighborhood $\omega$ of $0$ in
$\Lambda$. In \cite{Za1} we gave a survey (limited essentially to
the characteristic 2 case) on the $\sigma^+$-cellular automaton
with kernel the constant function 1 in $\omega$. In the present
paper we deal with general convolution operators over a field of
characteristic $p>0$. Our approach is based on the harmonic
analysis. We address the problem of determining the spectrum of a
convolution operator in the spaces of pluri-periodic functions on
$\Lambda$. This is equivalent to the problem of counting points on
the associate algebraic hypersurface in an algebraic torus
according to their torsion multi-orders. These problems lead to a
version of the Chebyshev-Dickson polynomials parameterized this
time by the set of all finite index sublattices of $\Lambda$ and
not by the naturals as in the classical case. It happens that the
divisibility property of the classical Chebyshev-Dickson
polynomials holds in this more general setting.
\end{abstract}

\maketitle

{\footnotesize
 MP: {\it - Do you yourself perceive a fundamental
difference between pure and applied mathematics$?$}

 Stanislaw Ulam: {\it - I really don't. I think it's a
question of language, and perhaps habits.}}

{\footnotesize \tableofcontents}

\section*{Introduction}
These notes are inspired by the theory of linear cellular
automata. Such an automaton on the integer lattice $\Z^s$ can be
viewed as a discrete dynamical system generated by a convolution
operator $f\longmapsto \Delta_a f=f*a$, acting on functions
$f:\Z^s\to K$ with values in a Galois field $K=\GF(p)$. Usually
the kernel $a$ of $\Delta_a$ is concentrated in the nearest
neighborhood of $0\in\Z^s$. We are interested more generally in
systems of convolution equations \be\label{csy}\Delta_{a_j}
f=0,\qquad j=1,\ldots,t\,\ee with kernel $\bar a=(a_1,\ldots,a_t)$
of bounded (and so finite) support. We address the following
questions.

\smallskip

{\bf Problem.} {\it Describe the set of all possible pluri-periods
of the pluri-periodic solutions of (\ref{csy}). More precisely,
given a pluri-period $\bar n\in\N^s$ compute the spectral
multiplicities of (\ref{csy}) on the space of all $\bar
n$-periodic functions on $\Z^s$ and the dimension of the
corresponding kernel $\ker \Delta_{\bar a}$.}

\smallskip

At present these problems seem to be out of reach. We provide
however different interpretations that could be useful in future
approaches.

Counting pluri-periods amounts to counting points on the
associated affine algebraic variety $\Sigma_{\bar a}$ (called
symbolic variety) according to their multi-orders. The symbolic
variety $\Sigma_{\bar a}$ is a subvariety in the algebraic torus
$(\bar K^\times)^s$, where $\bar K$ stands for the algebraic
closure of $K$. The multiplicative group $\bar K^\times$ being a
torsion group, the torus is covered by the finite subgroups. We
are interested in the distribution of points on $\Sigma_{\bar a}$
according to the filtration of $(\bar K^\times)^s$ by finite
subgroups.

The spectral multiplicities which appear in the above problem can
be described via Chebyshev-Dickson polynomial systems. Such a
system associates a degree $d$ polynomial in $K[\lambda]$ to any
sublattice of $\Z^s$ of index $d$, namely the characteristic
polynomial of (\ref{csy}) in the corresponding function space. The
classical Chebyshev-Dickson polynomials appear in the simplest
case where $s=1$. In Theorem \ref{divisthm} we establish the
divisibility property for Chebyshev-Dickson systems. Besides, we
give in Proposition \ref{syb3} a description of these systems via
iterated resultants.

Using the harmonic analysis we interpret the points of
$\Sigma_{\bar a}$ as $\bar a$-harmonic lattice characters; see
Theorem \ref{manyop} below. Here '$\bar a$-harmonic` simply means
'satisfying (\ref{csy})`. In the classical case, for a solution of
(\ref{csy}) the value in a lattice point is a sum of its values
over the neighbor points\footnote{In positive characteristic, one
has to replace averaging by summation.}; this explains our
terminology.

Resuming, in these notes we explore interplay between periods of
solutions of a system of convolution equations on a lattice, on
one hand, torsion orders of points on the associate symbolic
variety, on the other hand, and harmonic characters. Let us
develop  along these lines in more detail.

\bsit\label{intro0} {\it $\sigma^+$-automaton.} In \cite{Za1} we
gave a survey on the $\sigma^+$-automata on rectangular and toric
grids. Let us recall the setup. On the integral lattice
$\Lambda=\Z^s$ we consider the following function $a^+$ with
values in the binary Galois field $\GF(2)$: \be\label{starf}
a^+=\delta_0+\sum_{i=1}^s (\delta_{e_i}+\delta_{-e_i})\,,\ee where
$e_1,\ldots,e_s$ stands for the canonical lattice basis. We let
$\D_{a^+}$ denote the convolution operator $f\longmapsto f*a^+$
acting on binary functions $f:\Lambda\to\GF(2)$. It generates a
discrete dynamical system on $\Lambda$ called a
$\sigma^+$-automaton studied e.g., in \cite{MOW}, \cite{Su},
\cite{GKW}, \cite{BR}, \cite{SB}, \cite{HMP}; see further
references in \cite{Za1}.\esit

\bsit\label{intro1} {\it  $\sigma^+$-game.} The
$\sigma^+$-automaton on the plane lattice $\Lambda=\Z^2$ is
related to the solitaire game 'Lights Out`, also called a
$\sigma^+$-game. Let us describe the game. Suppose that the
offices in a department, which will be our table of game,
correspond to the vertices of a grid $P_{m,n}=L_m\times L_n$,
where $L_m$ stands for the linear graph with $m$ vertices. Suppose
also that the interrupters are synchronized in such an uncommon
way that turning off or on in one room changes automatically to
the opposite the states in all rooms neighbors through a wall. The
question arises whether the last person leaving the department can
always manage to turn all the lights off.

It is possible to reduce this problem to an analogous one for the
toric grid $\T_{m',n'}= C_{m'}\times C_{n'}$, where
$m'=m+1,\,n'=n+1$ and $C_m$ stands for the circular graph with $m$
vertices. We let $\cF=\cF(\T_{m,n},\GF(2))$ be the function space
on the torus equipped with the standard bilinear form $\langle
\cdot, \cdot\rangle $. The move at a vertex $v$ of $\T_{m,n}$ in
the $\sigma^+$-game, applied to a function (a 'pattern`)
$f\in\cF(\T_{m,n},\GF(2))$, consists in the addition
$$f\longmapsto f+a_v^+\mod 2\,,$$
where $a_v^+(u)=a^+(u+v)$ is the shifted star function
(\ref{starf}) centered at $v\in\T_{m,n}$. Thus the $\sigma^+$-game
on the torus $\T_{m,n}$ is winning starting with the initial
pattern $f_0$ if and only if $f_0$ can be  decomposed into a sum
of shifts of the star function $a^+$.

The linear invariants of the $\sigma^+$-game form a subspace
$\cH\subseteq \cF$ orthogonal to all shifts $a_v^+$, $v\in
\T_{m,n}$. Indeed
$$h\in
\cH\quad\Longleftrightarrow\quad \langle h, f+a_v^+\rangle\equiv
\langle h, f\rangle \mod 2\qquad \forall v\in
\T_{m,n},\,\,\,\forall f\in \cF\,.$$ Moreover the initial pattern
$f_0$ is winning if and only if $f_0\in \cH^\bot$. The functions
$h\in\cH$ are called harmonic \cite{Za1}. This is justified  by
the following property: for any vertex $v$ of the grid $\T_{m,n}$,
the value $h(v)$ is the sum modulo 2 of the values of $h$ over the
neighbors of $v$ in $\T_{m,n}$. Actually $\cH=\ker (\D_{a^+})$.
Thus the $\sigma^+$-game on a toric grid $\T_{m,n}$ is winning for
any initial pattern if and only if $ 0\notin \spec (\D_{a^+})$.
The latter is known to be equivalent to the condition $\gcd
(T_m,T_n^+)=1$, where $T_m$ stands for the classical $m$th
Chebyshev-Dickson polynomial over $\GF(2)$ and
$T_n^+(x)=T_n(x+1)$, see \cite[2.35]{Za1} and the references
therein. \esit

\bsit\label{intro35} This discussion leads to the following
problems.

\smallskip

\begin{enumerate}\item[$\bullet$]
{\it Determine the set of all winning toric grids $\T_{m,n}$.
Equivalently, determine the set of all pairs $(m,n)$ such that the
polynomials $T_m$ and $T_n^+$ are coprime. Or, which is
complementary, determine the  set of all toric grids $\T_{m,n}$
admitting a nonzero binary harmonic function. \item[$\bullet$]
Given $(m,n)\in \N^2$ compute the dimension $d(m,n)$ of the
subspace $\cH$ of all harmonic functions on $\T_{m,n}$ or,
equivalently, the dimension of the subspace $\cH^\bot$ of all
winning patterns. }
\end{enumerate}

\smallskip

For $m,n$ odd we provide several different interpretations of
$d(m,n)$. In particular we will show that, over the algebraic
closure $\bar K$ of the base field $K=\GF(2)$, there is an
orthonormal basis of $\cH\otimes\bar K$ consisting of harmonic
characters on $\T_{m,n}$ with values in the multiplicative group
$\bar K^\times$. An initial pattern $f_0$ on $\T_{m,n}$ is winning
if and only if $f_0$ is orthogonal to all harmonic characters on
$\T_{m,n}$. The latter ones are in one to one correspondence with
the $(m,n)$-bi-torsion points on the symbolic hypersurface. In our
case the symbolic hypersurface is the elliptic cubic in the torus
$({\bar K^\times})^2=(\A^1_{\bar K}\setminus\{0\})^2$ with
equation \be\label{cub} x+x^{-1}+y+y^{-1}+1=0\,.\ee Thus to
determine all toric grids $\T_{m,n}$ admitting a nonzero binary
harmonic function is the same as to determine all bi-torsion
orders of points on the cubic (\ref{cub}), see \cite{Za1}. \esit

\bsit\label{intro3} {\it Linear cellular automata on abelian
groups.} In the present paper we consider similar problems for
general linear cellular automata on abelian groups. Recall that
the theory of cellular automata aims, in particular, to provide a
model for inter-cellular or inter-molecular interactions. One can
regard a linear cellular automaton on a group, or rather on the
Caley graph of a group, as a discrete dynamical system generated
by a convolution operator with kernel concentrated in a nearest
neighborhood of the neutral element \cite{MOW}.

In more detail, suppose we are given a collection (a colony) of
'cells` placed at the vertices of a locally finite graph $\G$.
This determines the relation 'neighbors` for cells; neighbors can
interact. Each cell can be in one of $n$ cyclically ordered
states; thus the state of the whole collection at a moment $t$ is
codified by a function $f_t:\G\to\Z/n\Z$. In the subsequent
portions of time, the cells simultaneously change their states.
According to a certain local rule, the new state of a cell depends
on the previous states of the given cell and of all its neighbors.

To define a cellular automaton, say, $\sigma$ on $\G$ means to fix
at each vertex $v$ of $\G$ such a local rule, which does not
depend on $t$. Such a collection of local rules determines a
discrete dynamical system $\sigma:f_t\longmapsto f_{t+1}$. Usually
the edges $[v_0,v_1],\ldots,[v_0,v_s]$ at $v_0$ are ordered. So
the local rule at $v_0$  is a function $\phi_{v_0}:
(\Z/n\Z)^{s+1}\to \Z/n\Z$, and \be\label{eveq}
f_{t+1}(v_0)=\phi_{v_0} \left(f_t(v_0),
f_t(v_1),\ldots,f_t(v_s)\right)\,.\ee

In the case where the graph $\G$ is homogeneous under a group
action on $\G$, it is natural to assume that the family of local
rules is as well homogeneous.
Consider for  instance  the Caley graph $\G$ of a finitely
generated group $G$ with a generating set $\{g_1,\ldots,g_s\}$.
Given a local rule $\phi_e$ for the neutral element $e\in G$, we
can define $\phi_g$ for any vertex $g\in G$ as the shift of
$\phi_e$ by $g$.

For an additive cellular automaton the local rule $\phi_e$ is a
linear function. Consequently such an automaton is generated by a
convolution operator $\D_a:f\longmapsto f*a$ on $G$ with kernel
$a$ supported in the nearest neighborhood of $e$. This kernel is
just the coefficient function of $\phi_e$. The evolution equation
(\ref{eveq}) can be written in this case as a heat equation
$$\partial_t
(f_t):=f_{t+1}-f_t=\D_{a'}(f_t),\qquad\mbox{where}\qquad
a'=a-\delta_e\,.$$

Here we restrict to additive cellular automata on lattices or
toric grids, viewed as the Caley graphs of finitely generated free
abelian groups and finite abelian groups, respectively. In
contrast with the classical setting, we allow distant
interactions. So we deal with general convolution operators. \esit

\bsit\label{intro4} {\it Convolution operators on lattices and
Chebyshev-Dickson systems.} For a field $K$ of characteristic
$p>0$ and for a group $G$ we let $\cF(G, K)$ and $\cF^0(G, K)$
denote the vector space of all functions $f:G\to K$, of all those
with finite support, respectively. We consider the convolution
$$*:\cF(G, K)\times\cF^0(G, K)\ni (f,a) \longmapsto f*a\in\cF(G,
K)\,,$$ where
$$f*a(g)=\sum_{h\in G} f(h) a(h^{-1}g)\qquad \forall g\in G\,.$$
Fixing $a$ we get the convolution operator $\D_a: f\longmapsto
f*a$ acting on the space $\cF(G, K)$. All such operators form a
$K$-algebra $\Conv_K (G)$ with
$\D_{a_1}\circ\D_{a_2}=\D_{a_2*a_1}$. \esit

\bsit\label{intro5} For a subgroup $H\subseteq G$ we let
$\D_a\vert H$ denote the restriction of $\D_a$ to the subspace
$\cF_{H}(G, K)\subseteq \cF(G, K)$ of all $H$-periodic functions.
Clearly $\cF_{H}(G, K)$ is of finite dimension whenever $H$ is of
finite index in $G$, and $\dim\cF_{H}(G,K)=[G\,:\,H]$.

In the sequel $G=\Lambda$ will be a lattice i.e., a free abelian
group of finite rank. We let $\cL$ denote the set of all finite
index sublattices in $\Lambda$. Ordered by inclusion, $\cL$ can be
regarded as an ordered graph. Given a function $a\in
\cF^0(\Lambda, K)$ with finite support, we consider the spectra
and the spectral multiplicities of $\D_a\vert \Lambda'$,
$\Lambda'\in\cL $, in the algebraic closure $\bar K$ of $K$. In
particular we consider the function \be\label{funct}
d(a,\Lambda')=\dim \ker\, (\D_a\vert \Lambda'), \qquad
\Lambda'\in\cL \,.\ee The family of characteristic polynomials
$$\charpol_{a,\Lambda'}=\charpol (\D_a\vert \Lambda'),
\qquad \Lambda'\in\cL,\,$$ will be called a Chebyshev-Dickson
system. Recall that the $n$th Dickson polynomial $D_n(x,\alpha)$
over a finite field $F$ is the unique polynomial verifying the
identity $D_n(x+\alpha/x,\alpha)=x^n+\alpha^n/x^n$, where
$\alpha\in F$. Whereas for $F=\GF(2)$, the $n$th Chebyshev-Dickson
polynomial of the first kind is $T_n(x)=D_n(x,1)$. We recover the
latter one as $T_n=\charpol_{a,\Lambda'}$ when taking $K=\GF(2)$,
$a=a^+-\delta_e$, $\Lambda=\Z$ and $\Lambda'=n\Z$.

The classical Chebyshev-Dickson system $(T_n)$ possesses a number
of interesting properties. It forms a commutative semigroup under
composition that is, $T_n\circ T_m=T_{mn}$. Furthermore $T_m$
divides $T_n$ if $m\mid n$, moreover, $\gcd (T_m,T_n)=T_{\gcd
(m,n)}$, etc. The composition property is not stable under shifts
in the argument, and so does not hold in our more general setting.
However, the divisibility property does hold. Namely we have the
following

\bthm\label{divisthm} $\charpol_{a,\Lambda''}$ divides
$\charpol_{a,\Lambda'}$ whenever $\Lambda'\subset \Lambda''$.
Moreover, if the index of $\Lambda'$ in $\Lambda''$ equals
$p^\alpha$ then
$\charpol_{a,\Lambda'}=(\charpol_{a,\Lambda''})^{p^\alpha}$. \ethm

The second assertion allows to restrict to the subgraph
$\cL^0\subseteq \cL$ of all sublattices $\Lambda'\subseteq\Lambda$
with indices coprime with $p$. For a sublattice
$\Lambda'\in\cL^0$, the dimension $d(a,\Lambda')$ of the kernel of
$\D_a$ equals the multiplicity of the zero root of the polynomial
$\charpol_{a,\Lambda'}$. \esit

\bsit\label{intro6} {\it Systems of convolution equations.} We let
$K=\GF(p)$ be the Galois field of characteristic $p>0$, $\bar K$
be the algebraic closure of $K$ and $\bar K^\times$ be the
multiplicative group of $\bar K$. We fix a cortege $\bar
a=(a_1,\ldots,a_t)$ consisting of functions $a_j:\Lambda\to\bar K$
 with bounded supports.

Given a lattice $\Lambda$ we consider the system of convolution
equations \be\label{sys0}
\D_{a_j}(f):=f*a_j=0,\,\,\,j=1,\ldots,t,\qquad\mbox{where}\quad
f:\Lambda\to\bar K\,.\ee We let $d(\bar a,\bar n)$ denote the
number of linearly independent $\bar n$-periodic solutions of
(\ref{sys0}). We call these solutions $\bar a$-harmonic.

We let
$$\charpol_{\bar a, \Lambda'}=
\gcd\, \left(\charpol_{ a_j,
\Lambda'}\,:\,j=1,\ldots,t\right)\,,$$
$$\ker (\D_{\bar a}\vert
\Lambda')=\bigcap_{j=1}^t \ker\left( \D_{a_j}\vert \Lambda'\right)
\qquad\mbox{and}\qquad d(\bar a,\Lambda')=\dim \ker(\D_{\bar
a}\vert \Lambda') \,.$$ Thus $\charpol_{\bar a,
\Lambda'}(\lambda)=0$ if and only if there exists a nonzero
$\Lambda'$-periodic eigenfunction $f\in\cF_{\Lambda'}(\Lambda,\bar
K)$ of $\D_{\bar a}$ with $$\D_{a_j} (f)=\lambda\cdot
f\quad\forall j=1,\ldots,t\,.$$ The set of zeros of
$\charpol_{\bar a, \Lambda'}$ counted with multiplicities is
called the spectrum of $\D_{\bar a}\vert \Lambda'$, and is denoted
by $\spec\, (\D_{\bar a}\vert \Lambda')$. The set of spectra forms
a graph $\Xi$ ordered by inclusion. Due to the divisibility
property in Chebyshev-Dickson systems, the map $\spec:\cL\to \Xi$
of ordered graphs is monotonous. \esit

\bsit\label{intro7} {\it Symbolic variety.} Given a basis
$\cV=(v_1,\ldots,v_s)$ of $\Lambda$ we can identify $\Lambda$ with
$\Z^s$. To a function $a:\Lambda=\Z^s\to \bar K$ with bounded
support we associate the Laurent polynomial
$$\sigma_{a}=
\sum_{u=(u_1,\ldots,u_s)\in\Z^s} a(u)x^{-u}\in \bar
K[x_1,x_1^{-1},\ldots, x_s,x_s^{-1}],\,\,\,j=1,\ldots,t.$$ A
cortege $\bar a=(a_1,\ldots,a_n)$ determines an affine algebraic
variety
$$\Sigma_{\bar a}=\{\sigma_{a_j}=0\,:\,j=1,\ldots,t\}\,$$
in the torus $(\bar K^\times)^s$. We call $\Sigma_{\bar a}$ the
symbolic variety associated with the system (\ref{sys0}).

The logarithm of the Weil zeta function counts the points on
$\Sigma_{\bar a}$ over the Galois fields $\GF(q)$, where $q=p^r$,
$r\ge 0$. Whereas our purpose is to count, for every multi-index
$\bar n=(n_1,\ldots,n_s)\in\N^s$ with all $n_i$ coprime to $p$,
the number $$d_{\bar a} (\bar n)=\card\, (\Sigma_{{\bar a}, \bar
n})\,$$ of $\bar n$-torsion points on the symbolic variety
$\Sigma_{\bar a}$, where
$$\Sigma_{{\bar a}, \bar n}=\{\xi=(\xi_1,\ldots,\xi_s)
\in\Sigma_{\bar a} \,:\, \xi_i^{n_i}=1,\,\,i=1,\ldots,s\}\,.$$ Due
to Theorem \ref{manyop}(a) below this same quantity arises in the
spectral problem:
$$d_{\bar a} (\bar n)=d(\bar a,\bar n)\,.$$  \esit

\bsit\label{intro8} {\it Harmonic characters.} We let $\Char
(\Lambda,\bar K^\times)$ denote the set of all $\bar
K^\times$-valued characters of $\Lambda$, that is of all
homomorphisms $\chi:\Lambda\to\bar K^\times$. A character $\chi$
is called $\bar a$-harmonic if the function $\chi:\Lambda\to\bar
K$ is; $\Char_{\bar a-{\rm harm}} (\Lambda,\bar K^\times)$ stands
for the set of all $\bar a$-harmonic characters of $\Lambda$.

We denote by $\NC$ the set of all naturals coprime with $p$. Given
a multi-index $\bar n\in\NC^s$ we consider the finite subgroup of
the torus $(\bar K^\times)^s$
$$ \mu_{\bar
n}:=\{\xi=(\xi_1,\ldots,\xi_s)\in (\bar
K^\times)^s\,:\,\xi_i^{n_i}=1,\,\,\,i=1,\ldots,s\}\,.$$ Fixing a
basis $\cV=(v_1,\ldots,v_s)$ of $\Lambda$ we consider the product
sublattices
$$\Lambda'=\Lambda_{\bar n,\cV}=\sum_{i=1}^s
n_i\Z v_i\subseteq\Lambda\,.$$ For such a sublattice
$\Lambda'\subseteq\Lambda$, the Fourier transform provides a
natural one to one correspondence between the set of all $\bar
a$-harmonic characters of the quotient group $G=\Lambda/\Lambda'$
and the set of points on the symbolic variety with multi-torsion
order dividing $\bar n$. Namely the following hold.\esit

\bthm\label{manyop}
\begin{enumerate}\item[(a)] For any sublattice $L'\in \cL^0$, the
subspace $\ker(\D_{\bar a}\vert \Lambda')$ possesses an
orthonormal basis of $\bar a$-harmonic characters. In particular
there are $d(\bar a,\Lambda')=\mult_{\lambda=0} \,(\charpol_{\bar
a, \Lambda'})$ such characters. \item[(b)] Fixing a basis $\cV$ of
$\Lambda$ provides a natural bijection $\Char (\Lambda,\bar
K^\times)\stackrel{\cong}{\longrightarrow} (\bar K^\times)^s$.
This bijection restricts to
$$\Char_{\bar a-{\rm harm}} (\Lambda,\bar K^\times)
\stackrel{\cong}{\longrightarrow} \Sigma_{\bar a}\subseteq (\bar
K^\times)^s\,.$$ Moreover $\forall \bar n\in\NC^s$ it further
restricts to
$$\Char_{\bar a-{\rm harm}}
(\Lambda/\Lambda_{\bar n,\cV},\bar K^\times)
\stackrel{\cong}{\longrightarrow} \Sigma_{\bar a,\bar
n}:=\Sigma_{\bar a}\cap\mu_{\bar n}\,.$$
\end{enumerate}\ethm

In Section 1 we prove the second part of Theorem \ref{divisthm},
see Corollary \ref{charp6}. The first parts of Theorems
\ref{divisthm} and \ref{manyop} are proven in Section 2, see
Corollary \ref{futra4} and Theorem \ref{divis}, respectively. We
also deduce an expression of polynomials in a Chebyshev-Dickson
system via iterated resultants, see Proposition \ref{syb3}.

In Section 3 we prove Theorem \ref{manyop}(b) (see Proposition
\ref{equa}). Besides, we provide in Theorem \ref{MFT} a dynamical
criterion for existence of a nonzero periodic $\bar a$-harmonic
function with a given pluri-period. Or what is the same, for
existence of a point on the symbolic variety\footnote{Which can be
an arbitrary affine algebraic variety in a torus.} with a given
multi-torsion order. In addition we provide a formula for the
orthogonal projection onto the space of all $\bar a$-harmonic
functions.

In the final Section 4 we study similar problems over finite
fields. Assuming $p$-freeness, in Theorem \ref{many} we show that
any $\bar a$-harmonic function with values in the original Galois
field $\GF(p)$ is a linear combination of traces of harmonic
characters.

In Section 5 we discuss translation invariant subspaces generated
by characters. The reader will also find here many concrete
examples. In the final Section 6 he will find a discussion
concerning the partnership graph, related to Zagier's theorem on
finiteness of the connected components of this graph in the case
of the $\sigma^+$-automaton  over the field $\GF(2)$ on a plane
lattice, see \cite{Za1}. We indicate here several open problems.

The author is grateful to Don Zagier for clarifying discussions,
in particular for the idea of processing in the present
generality. Maxim Kontsevich suggested that the function $d_{\bar
a} : \N^s\to \N$ should be studied via the technique of
statistical sums within the framework on the Ising model. Our
thanks also to Vladimir Berkovich for a kind assistance and to
Dmitri Piontkovski for performing computer simulations.

\section{Sylow $p$-subgroups and
Chebyshev-Dickson systems} The Dickson polynomials $D_n(x,a)$
($E_n(x,a)$, respectively) of the first (second) kind over a
finite field of characteristic $p>0$ can be characterized by the
identities
$$\mu^n_1+\mu^n_2=D_n(\mu_1+\mu_2,\mu_1\mu_2)\qquad \mbox{resp.}
\qquad\mu^{n+1}_1-\mu^{n+1}_2/(\mu_1-\mu_2)=E_n(\mu_1+\mu_2,\mu_1\mu_2).$$
They also satisfy the relations \cite{BZ} $$D_{p^\alpha
m}=(D_m)^{p^\alpha}\qquad\mbox{resp.}\qquad E_{p^\alpha
m}=(E_m)^{p^\alpha}\,.$$ In this section we show that analogous
relations hold for any Chebyshev-Dickson polynomial system
$\charpol_a : \cL\to K[x]$  (see Corollary \ref{charp6}). Here $K$
denotes a field of characteristic $p>0$, $\Lambda$ a lattice,
$a\in \cF^0(\Lambda, K)$ a $K$-valued function on $\Lambda$ with
bounded support and $\cL$ the ordered graph of all finite index
sublattices $\Lambda'\subseteq\Lambda$, see \S\ref{intro5} in the
Introduction. This allows to recover $\charpol_a$ by its
restriction to the subset $\cL^0\subseteq \cL$ of all sublattices
with indices coprime to $p$.

For a finite group $G$ and a subset $A\subseteq G$ we let
$\delta_A=\sum_{u\in A} \delta_u$ denote the characteristic
function of $A$, where $\delta_u$ stands for the delta-function on
$G$ concentrated on $u\in G$. For a function $a=\sum_{u\in G}
a(u)\delta_u$ on $G$ we let
$$|a\vert A|=\sum_{u\in A} a(u),\quad
\charpol_{a,G}=\det (\D_a-\lambda\cdot 1)\quad\mbox{and}\quad
d(a,G)=\dim\ker (\D_a)\,.$$

The following Pushforward Lemma is simple, so we leave the proof
to the reader.

\blem\label{strf} For a normal subgroup $H\subseteq G$ and for any
$a\in\cF^0(G,K)$, the function \be\label{aver1} a_*(v+H)=|a\vert
(v+H)|=\sum_{v'\in H} a(v+v')\,\ee is a unique function in
$\cF^0(G/H,K)$ satisfying $a_*\circ \pi=a*\delta_H$, where $\pi:
G\twoheadrightarrow G/H$ is the canonical surjection. Moreover,
there is a commutative diagram
\begin{diagram} \cF_{H}(G,K)&\rTo^{\D_a}&\cF_{H}(G,K)\\
\dTo_\cong & &\dTo_\cong \\
\cF (G/H,K)&\rTo^{\D_{a_*}}&\cF (G/H,K)\,,
\end{diagram} where $$\cF_{H}(G,K)=\{f\in\cF(G,K)\,:\,\tau_h (f)
=f\quad\forall h\in H\}\quad\mbox{and}\quad \tau_h
(f)(g)=f(gh)\,.$$\elem

The convolution algebra $\Conv_K\, (G)$ consists of all operators
on the space $\cF^0(L, K)$ commuting with shifts. Moreover the
shifts $(\tau_u\,:\,u\in G)$ generate $\Conv_K\, (G)$ as a
$K$-vector space. Indeed $\forall a\in\cF^0(G,K)$ one has
\be\label{decoshi} \D_a=\sum_{g\in G} a(g) \tau_{g^{-1}}\,.\ee
Notice that $\tau_{g^{-1}}=\D_{\delta_{g}}$ and
$a=\D_a(\delta_e)$, where $e\in G$ denotes the neutral element.

For an endomorphism $A\in\End (\A^n_K)$ we let $A=S_A+N_A$ be the
Jordan decomposition, where $S_A$ is semisimple, $N_A$ nilpotent
and $S_A,N_A$ commute. It is defined over the algebraic closure
$\bar K$ of $K$.

\bprop\label{charp3} Let $G=F\times H$ be a direct product of two
abelian groups, where $H=\bigoplus_{i=1}^n \Z/p^{r_i}\Z$. Then for
any $a\in\cF^0(G,K)$ the following hold.
\begin{enumerate} \item[(a)]
$S_{\D_a}=S_{\D_{a_*}}\otimes 1_H$, where $a_*\in\cF(F, K)$ is as
in (\ref{aver1}) above. \item[(b)]
$\charpol_{a,G}=(\charpol_{a_*,F})^{\ord \,H}$.
\end{enumerate}
Consequently there exists a nonzero $a$-harmonic function on $G$
if and only if there is a nonzero $a_*$-harmonic function on
$F$.\eprop

%



\bproof (a) follows by induction on $n$, once it is established
for $n=1$. Letting $H=\Z/p^r\Z$ we will show that \be\label{pr}
\D_a^{p^r}=\D_{a_*}^{p^r}\otimes 1_H\,.\ee To this end,
decomposing $u\in G$ as $u=u'+u''$ with $u'\in F$ and $u''\in H$,
we obtain
$$\D_a^{p^r}=\sum_{u\in G} [a(u)]^{p^r} (\tau_{-u})^{p^r}=
\sum_{u'\in F} \left(\sum_{u''\in H}
[a(u'+u'')]^{p^r}\right)(\tau_{-u'})^{p^r}$$
$$=\left(\sum_{u'\in F}
a_*(u')\tau_{-u'}\right)^{p^r}\otimes 1_H=\D_{a_*}^{p^r}\otimes
1_H\,.$$ This proves (\ref{pr}). By virtue of (\ref{pr}) we get
$$S_{\D_a}^{p^r}+N_{\D_a}^{p^r}=S_{\D_{a_*}}^{p^r}\otimes 1_H
+N_{\D_{a_*}}^{p^r}\otimes 1_H\,.$$ By the uniqueness of the
Jordan decomposition we have
$S_{\D_a}^{p^r}=S_{\D_{a_*}}^{p^r}\otimes 1_H$ and so (a) follows.
Now (b) is immediate from (a). \eproof

From Proposition \ref{charp3}
we deduce the following corollaries.

\bcor\label{charp5} If $G=\bigoplus_{i=1}^k \Z/p^{r_i}\Z$, where
$p=\cchar K$, then
$$\charpol_{a,G}=(x-|a|)^{\ord\, G}\,.$$ In particular there exists an
$a$-harmonic function on $G$ if and only if the constant function
$1$ on $G$ is $a$-harmonic, if and only if $|a|=0$. \ecor

\brem\label{ctrex} Letting $K=\GF(2)$, $G=\Z/4\Z$ and
$a=a^+=\delta_{\bar 0}+\delta_{\bar 1}+\delta_{-\bar
1}\in\cF(G,K)$ we obtain $\charpol_{a,G}=(x+1)^4$. Hence the
algebraic multiplicity of the spectral value $\lambda=1$ of
$\Delta_a$ equals 4. Although $S_{\Delta_a}=1_G$, the nilpotent
part $N_{\Delta_a}$ is present and $\dim \ker (\Delta_a+1_G)=2$.
\erem

Given $\bar a=(a_1,\ldots a_t)\in (\cF^0(G, K))^t$ we let as
before
$$\ker\,(\D_{\bar a})=\bigcap_{j=1}^t \ker\,(\D_{a_j}),\qquad d(\bar a,
G)=\dim \ker\,(\D_{\bar a})$$ and
$$\charpol_{\bar a,G}=\charpol (\D_{\bar a})=\gcd \left(\charpol
(\D_{a_j})\,:\,j=1,\ldots,t\right)\,.$$ For a subgroup $H\subseteq
G$ we let
$$\charpol_{\bar a_*,G/H}=\charpol (\D_{\bar a}\vert
\cF_H(G, K))\,,$$ where $\bar
a_*=((a_1)_*,\ldots,(a_n)_*)\in\cF(G/H,  K)$.

\bcor\label{charp6} Let $\Lambda$ be a lattice and
$\Lambda'\subseteq \Lambda$ a sublattice of index $p^\alpha q$,
where $q\not\equiv 0 \mod p$. Then there exists a unique
intermediate sublattice $\Lambda''$ of index $q$ in $\Lambda$,
where $\Lambda'\subseteq \Lambda''\subseteq \Lambda$. Moreover
$\forall \bar a\in (\cF^0(\Lambda, K))^t$ one has \be\label{chpo}
\charpol_{\bar a,\Lambda'}=(\charpol_{\bar a,\Lambda''})^{p^\alpha
} \,.\ee  \ecor

\bproof It is enough to show (\ref{chpo}) for $t=1$; then  it
follows easily for any $t\ge 1$. So letting $t=1$ and $a_1=a$ we
decompose
$$G=\Lambda/\Lambda'=F\oplus G(p)\,,$$ where $G(p)$ is the Sylow
$p$-subgroup of $G$ and $\ord\, F=q$. We let
$\Lambda''=\pi'^{-1}(G(p))$, where $\pi':\Lambda\twoheadrightarrow
G$, so that $\Lambda/\Lambda''\cong F$. Due to the Pushforward
Lemma \ref{strf}, \be\label{chpo1} \charpol_{a,\Lambda'}=
\charpol_{\pi'_*a,G}\quad\mbox{ and} \quad
\charpol_{a,\Lambda''}=\charpol_{\pi''_*a,F}\,,\ee where
$\pi'':\Lambda\twoheadrightarrow F$. Now (\ref{chpo}) follows from
(\ref{chpo1}) in view of Proposition \ref{charp3}(b).  \eproof

\section{Chebyshev-Dickson systems in the $p$-free case}
\subsection{Naive Fourier transform
on convolution algebras} For a finite group $G$ there are natural
isomorphisms
$$\left(\cF(G,K),*\right)\stackrel{\varphi}{\longrightarrow}
\Conv_K\, (G)\stackrel{\psi}{\longrightarrow} K[G]\,,$$ where
$\varphi:a\longmapsto \D_a$, $\Conv_K\, (G)$ is the convolution
algebra and $K[G]$ is the group algebra of $G$ over $K$. For
instance \cite{MOW}, the group ring of a finite abelian group
$$G=\bigoplus_{i=1}^s
\Z/n_i\Z$$ is the truncated polynomial ring
$$K[G]=\bigotimes_{i=1}^s
K[x_i]/(x^{n_i}_i-1)\,.$$ The ideals of $\cF(G,K)$ are called
convolution ideals. In particular, for any subgroup $H\subseteq
G$, the translation invariant subspace
$$\cF_{H}(G,K)=\{f\in\cF(G,K)\,:\,\tau_h (f)
=f\quad\forall h\in H\}\,$$ is a convolution ideal.

 The composition $\psi\circ\varphi:\cF(G,K)\to K[G],\quad a\longmapsto
 \tilde a$, provides a naive Fourier transform, which sends
$\D_a$ to the operator of multiplication by $\tilde a$ in $K[G]$
and $\ker (\D_a)$ to the annihilator ideal $\Ann (\tilde a)$,
where $(\tilde a)\subseteq K[G]$ is the principal ideal generated
by $\tilde a$. Thus $G$ possesses a nonzero $a$-harmonic function
if and only if $\tilde a\in K[G]$ is a zero divisor.

The next result follows immediately from the Burnside Theorem.
Alternatively it can be deduced using the Fourier transform, see
below.

\blem\label{futra3} For any finite abelian  group $G$ of order
coprime to $p$ the following hold.
\begin{enumerate}\item[(a)]
$\cF(G,\bar K)$ admits a decomposition into a direct sum of
one-dimensional $\Conv_{\bar K}(G)$-submodules generated by
characters:
$$
\cF(G,\bar K)=\bigoplus_{g^\vee\in G^\vee} (g^\vee)\,.$$

\item[(b)] Any convolution ideal $I\subseteq \cF(G,\bar K)$ is
principal, generated by the sum of characters contained in $I$.
Furthermore there is a decomposition
$$\cF(G,\bar K)=I\oplus \Ann (I)\,.$$
In particular, for any subgroup $H\subseteq G$  one has
$$\cF(G,\bar K)=\cF_H(G,\bar K)\oplus \Ann
(\cF_H(G,\bar K))\,,$$ where
$$\cF_H(G,\bar K)
=\sum_{H\subseteq\ker (g^\vee)} \left(g^\vee\right)\,.$$
\end{enumerate}\elem

\subsection{Dual group and Fourier transform}
In the sequel we let $K=\GF(p^r)$. Thus the multiplicative group
$\bar K^\times$
is a torsion group, with torsion orders coprime with $p$. We let
$G$  be a finite abelian group of order coprime with $p$, and
$\NC$ be the set of all positive integers coprime to $p$.

The field $\bar K$ contains all roots of unity with orders
dividing $\ord \,G$. Hence the characters of $G$ can be realized
as homomorphisms $G\to \bar K^\times$.
This defines a natural embedding $G^\vee\hookrightarrow\cF(G,\bar
K)$.

The Fourier transform $F: \cF(G, \bar K)\to \cF (G^\vee, \bar K)$
is defined as usual \cite{Nic} via \be\label{FuTr} F:f\longmapsto
\widehat f,\quad\mbox{where}\quad \widehat f (g^\vee)=\sum_{g\in
G} f(g)g^\vee (g),\quad
 g^\vee \in G^\vee\,.\ee Its inverse
$F^{-1}: \cF (G^\vee, \bar K)\to \cF (G, \bar K)$ can be defined
via $$ F^{-1}:\varphi\longmapsto  \widehat
\varphi,\quad\mbox{where}\quad \widehat\varphi (g)=\frac{1}{\ord
\,(G)}\sum_{g^\vee\in G^\vee} \varphi(g^\vee)g^\vee (g^{-1}),\quad
g \in G\,.$$ With this notation $\Hat{\Hat f}=f$ and $\Hat{\Hat
\varphi}=\varphi$, which does not lead to a confusion as we never
exploit the Fourier transform on the dual group $G^\vee$.

Up to constant factors, both $F$ and $F^{-1}$ send
$\delta$-functions to characters and vice versa. Namely $\forall
g\in G, \, \forall g^\vee\in G^\vee$,
$$\widehat{\delta_g}=g,\quad\widehat{g}
=\delta_g\quad\mbox{and}\quad \widehat{\delta_G} =\sum_{g\in G}
\widehat{\delta_g}= \ord \,(G)\delta_{e^\vee}\,,$$ respectively,
$$\ord\, (G)\widehat{\delta_{g^\vee}}
=(g^\vee)^{-1},\quad\widehat{g^\vee}=\ord \,(G)
\delta_{(g^\vee)^{-1}}\quad\mbox{and}\quad
\widehat{\delta_{G^\vee}} =\delta_e\,.$$

Furthermore $F$ sends the convolution in the ring $\cF(G, \bar K)$
into the pointwise multiplication on $\cF(G^\vee, \bar K)$ giving
an isomorphism of $\bar K$-algebras. The convolution operator
$\D_a$ is sent to the operator of multiplication by $\Hat a$. The
Fourier transform of a character being proportional to a
delta-function, any character $g^\vee\in G^\vee$, up to a scalar
factor, is a convolution idempotent, and any convolution
idempotent of $(\cF(G,\bar K),*)$ is proportional to a sum of
characters.

 In the $\bar K$-vector space $\cF(G,\bar K)$ we
consider the non-degenerate symmetric bilinear form
 $$\langle f_1, f_2\rangle_1=\frac{1}{\ord \,(G)}\sum_{g\in G}
f_1(g)f_2(g^{-1})=\frac{1}{\ord \,(G)}f_1 * f_2 (e)\,.$$ Its
Fourier dual is  the following bilinear form in $\cF(G^\vee,\bar
K)$:
$$\langle \varphi_1, \varphi_2\rangle_2=\frac{1}{(\ord\, (G))^2}
\sum_{g^\vee\in G^\vee} \varphi_1(g^\vee)\varphi_2(g^\vee)\,.$$
Indeed we have $\langle \widehat {f_1}, \widehat
{f_2}\rangle_2=\langle f_1, f_2\rangle_1\,.$ The characters
$(g^\vee\,:\, g^\vee\in G^\vee)$ form an orthonormal basis in
$\cF(G,\bar K)$ with respect to the form $\langle \cdot,
\cdot\rangle_1$.

\subsection{Divisibility in Chebyshev-Dickson systems}
In this section $G$ denotes a finite abelian group of order
coprime to $p$. For a $t$-tuple $\bar a=(a_1,\ldots a_t)\in
(\cF^0(G,\bar K))^t$ we let as before $\ker\,(\D_{\bar
a})=\bigcap_{j=1}^t \ker\,(\D_{a_j})$ and
$$\charpol_{\bar a,G}=\charpol (\D_{\bar a})=\gcd \left(\charpol
(\D_{a_j})\,:\,j=1,\ldots,t\right)\,.$$ By the Pushforward Lemma
\ref{strf}, for a subgroup $H\subseteq G$ we have
$$\charpol_{\bar a_*,G/H}=\charpol (\D_{\bar a}\vert
\cF_H(G,\bar K))\,.$$  We let
$$V(a_j)=\Hat a_j^{-1}(0)\subseteq
G^\vee\quad\mbox{and}\quad  V(\bar a)=\bigcap_{j=1}^t
V(a_j)\subseteq G^\vee\,.$$ We also let $\Char_{\bar a-{\rm
harm}}(G,\bar K^\times)$ denote the set of all $\bar a$-harmonic
characters of $G$. The following corollary is straightforward from
Lemma \ref{futra3}.

\bcor\label{futra4}
\begin{enumerate}\item[(a)]
For any $a\in \cF^0(G,\bar K)$, the characters form an orthonormal
basis $\cF(G,\bar K)$ of eigenfunctions of $\D_a$, the matrix of
$\D_a$ in this basis is diagonal and so
$$\charpol (\D_a)=\prod_{g^\vee\in G^\vee}
\left(x-\Hat a (g^\vee)\right)\,.$$ Consequently $\quad\spec
(\D_a)= \Hat a (G^\vee)\subseteq \bar K,\qquad\ker
(\D_a)=\left(\widehat{\delta_{V(a)}}\right)$ and $$d(a, G)=\card\,
V(a)=\mult_0 \charpol (\D_a)\,.$$ \item[(b)] Similarly for any
$\bar a=(a_1,\ldots,a_t)\in(\cF^0(G,\bar K))^t$,
 the $\bar a$-harmonic characters form an
orthonormal basis in $\ker (\D_{\bar a})$,
$$\ker\,(\D_{\bar a})=
\left(\widehat{\delta_{V(\bar a)}}\right)\qquad\mbox{and}\quad
d(\bar a,G)= \card\, V(\bar a)\,.$$ Moreover there is a bijection
$V(\bar a)\cong \Char_{\bar a-{\rm harm}}(G,\bar K^\times)$.
Consequently the group $G$ admits a nonzero $\bar a$-harmonic
function if and only if it admits an $\bar a$-harmonic character.
\end{enumerate}\ecor

A convolution ideal $I\subseteq \cF(G,\bar K)$ and the annihilator
ideal $\Ann (I)$ being $\D_{\bar a}$-stable (see Lemma
\ref{futra3}), the Pushforward Lemma \ref{strf} yields the
following result.

\bprop\label{sex} For any convolution ideal $I\subseteq \cF(G,\bar
K)$, any subgroup $H\subseteq G$ and  any $\bar a\in (\cF^0(G,\bar
K))^t$ one has \be\label{sex0} \charpol (\D_{\bar a}\vert I)
\mid\charpol (\D_{\bar a})\quad\mbox{and}\quad\charpol_{\bar
a_*,G/H} \mid\charpol_{\bar a,G}\,.\ee \eprop

Now we can readily deduce the divisibility property in
Chebyshev-Dickson systems disregarding the assumption of
$p$-freeness.

\bthm\label{divis} For any  $\Lambda_1,\Lambda_2\in \cL$ we have:
$$\charpol_{\bar a,\Lambda_1+\Lambda_2}\mid
\gcd(\charpol_{\bar a,\Lambda_1}, \,\charpol_{\bar a,\Lambda_2})$$
and
$$\lcm(\charpol_{\bar a,\Lambda_1}, \,\charpol_{\bar a,\Lambda_2})
\mid\charpol_{\bar a,\Lambda_1\cap \Lambda_2}\,.$$ Consequently $
\charpol_{\bar a,\Lambda_2}\mid\charpol_{\bar a,\Lambda_1}$
whenever $\Lambda_1\subseteq \Lambda_2$. \ethm

\bproof It suffices to show the latter assertion, then the former
ones follow immediately. For a chain of finite index sublattices
$\Lambda'\subseteq \Lambda''\subseteq \Lambda$ we let
$G=\Lambda/\Lambda'$, $H=\Lambda''/\Lambda'$ so that
$G/H=\Lambda/\Lambda''$, and
$$\bar a'=\pi'_*\bar a\in\cF(G,\bar K),\qquad \bar a''=\pi''_*\bar
a\in\cF(G/H,\bar K)\,,$$ where $\pi:G\twoheadrightarrow G/H$,
$\pi':\Lambda\twoheadrightarrow  G$ and $\pi''=\pi\circ
\pi':\Lambda\twoheadrightarrow G/H$.

We assume first that the index of $\Lambda'$ in $\Lambda$ is
coprime with $p$ and so $\Lambda',\Lambda''\in\cL^0$. By the
Pushforward Lemma \ref{strf} we obtain
$$\charpol_{\bar a,\Lambda'}=\charpol_{\bar a', G}
\quad\mbox{and}\quad \charpol_{\bar a,\Lambda''} = \charpol_{\bar
a'', G/H}\,.$$ Hence by (\ref{sex0}) \be\label{sex1}\charpol_{\bar
a, \Lambda''} \mid \charpol_{\bar a, \Lambda'}\,.\ee

In the general case, assuming that $\Lambda_1\subseteq \Lambda_2$
we consider the decomposition $G_1=\Lambda/\Lambda_1=F\oplus
G_1(p)$, where $G_1(p)\subseteq G_1$ is the Sylow $p$-subgroup.
Letting $G_2=\Lambda_2/\Lambda_1\subseteq G_1$ we obtain $G_2(p)=
G_1(p)\cap G_2$. For the sublattices
$\Lambda_i''=\pi^{-1}(G_i(p))\subseteq \Lambda$, $i=1,2$, where
$\pi:\Lambda\twoheadrightarrow G_1$, we have $\Lambda_i''\supseteq
\Lambda_i$
 and $\Lambda_1'',\, \Lambda_2''\in\cL^0$.
 Since also $\Lambda_1''\subseteq \Lambda_2''$, by virtue of
 (\ref{sex1}) we obtain
 $\charpol_{a,\Lambda'_2}\mid\charpol_{a,\Lambda'_1}$.
 By (\ref{chpo})
 $\charpol_{a,\Lambda_i}=(\charpol_{a,\Lambda''_i})^{p^{\alpha_i}}$,
 $i=1,2$, where
 $\alpha_2\le\alpha_1$. Now the result follows.
\eproof

\brem\label{confun} Letting $\Lambda''=\Lambda$ we deduce that
$(x-|a|)\mid \charpol_{a,\Lambda'}$ $\forall
\Lambda'\in\cL,\,\,\forall a\in\cF^0(\Lambda,\bar K)$. We note
that the eigenspace in $\cF(\Lambda,\bar K)$, which corresponds to
the eigenvalue $|a|$ of $\D_a$, contains the constant function
$1$; cf. Corollary \ref{charp5}. \erem

\bexas\label{syb4} 1. Letting $G=G_1\times G_2$ and $a=a_1\otimes
a_2\in\cF(G,\bar K)$, where $a_i\in\cF(G_i,\bar K)$, $i=1,2$, we
obtain $$\charpol (\D_a) = \prod_{i,j} (x-\lambda_i\mu_j)\,,$$
where $\lambda_1,\ldots,\lambda_{\ord\, (G_1)}$ and
$\mu_1,\ldots,\mu_{\ord \,(G_2)}$ denote the eigenvalues of
$\D_{a_1}$ and $\D_{a_2}$, respectively.

2. Similarly, letting $a=a_1\otimes 1 \oplus 1\otimes
a_2\in\cF(G,\bar K)$, where again $G=G_1\times G_2$, we obtain
$$\charpol (\D_a) = \prod_{i,j} (x-(\lambda_i+\mu_j))\,.$$
The latter formula applies in particular to the graph Laplacians
with kernels $a_i^-=a^+_{G_i}-\delta_0$, $i=1,2$ and
$a^-=a^+_{G}-\delta_0$, respectively, where $a^+$ is the
star-function as in (\ref{starf}). In the characteristic 2 case
cf. Bacher's Lemma 2.10(a) in \cite{Za1}. \eexas

\subsection{Symbol of a convolution operator}
\bdefi\label{symbol}  Fixing a lattice $\Lambda$ of rank $s$, a
basis $\cV=(v_1,\ldots,v_s)$ of  $\Lambda$ and an $n$-tuple $\bar
n=(n_1,\ldots,n_s)$, where $n_i\in\N$, we let
$$
\Lambda_{\bar n, \cV}=\sum_{i=1}^s n_i\Z v_i\cong
\bigoplus_{i=1}^s n_i\Z$$ be the product sublattice of $\Lambda$
generated by $n_1v_1,\ldots,n_sv_s$. There is an isomorphism of
$K$-algebras
$$\sigma_{\cV} : \Conv_K(\Lambda) \stackrel{\cong}{\longrightarrow}
K[x_1,x_1^{-1},\ldots,x_s,x_s^{-1}],\qquad \D_a\longmapsto
\Symb_{a,\cV}\,,$$ which associates to a convolution operator
$\D_a$ on $\Lambda$ its $\cV$-symbol, that is the Laurent
polynomial in $s$ variables \be\label{symbol} \Symb_{a,\cV}=
\sum_{v=\sum_{i=1}^s \alpha_i e_i\in \Lambda} a(v) x^{-\alpha(v)}=
\sum_{v=\sum_{i=1}^s \alpha_i e_i\in \Lambda} a(v)x_1^{-\alpha_1}
\cdot\ldots\cdot x_s^{-\alpha_s}\,.\ee The inverse
$\sigma_{\cV}^{-1}$ is given by
$$x_i^{-1}\longmapsto
\tau_{v_i}\qquad\mbox{ and}\qquad x_i
\longmapsto\tau_{-v_i},\qquad i=1,\ldots,s\,.$$ The algebraic
hypersurface in the $s$-torus
\be\label{symbhyp}\Sigma_{a,\cV}=\Symb_{a,\cV}^{-1}(0)\subseteq
(K^\times )^s\,\ee associate with $\D_a$ will be called the
symbolic hypersurface. Similarly, for a sequence of convolution
operators $\D_{\bar a}=(\D_{a_j}\,:\,j=1,\ldots,t)$ its symbolic
variety is the affine subvariety in the torus
\be\label{symbvar}\Sigma_{\bar a,\cV}=
\bigcap_{j=1}^t\Symb_{a_j,\cV}^{-1}(0)\subseteq (K^\times
)^s\,.\ee \edefi

\bexa\label{four11} (see \cite{Za1}) For $K=\GF(2)$,
$\Lambda=\Z^2$, $\cV=(e_1,e_2)$ and $a=a^+$ the symbolic
hypersurface $\Sigma_{a^+,\cV}$ is the elliptic cubic (\ref{cub})
in $(\bar K^\times)^2$. Alternatively, this cubic can be given by
the equation
$$x^2y+xy^2+xy+x+y=0\,.$$
\eexa

For a finite abelian group $G=\Z_{\bar n}=\bigoplus \Z/n_i\Z$,
where $\bar n=(n_1,\ldots,n_s)\in\NC^s$, we let
$\cU=(e_1,\ldots,e_s)$ denote the standard base of $G$. We let
also
$$\mu_{\bar n}=\bigoplus_{i=1}^s
\mu_{n_i}\subseteq (\bar K^\times)^s\,,$$ where $\mu_n\subseteq
\bar K^\times$ stands for the cyclic group of $n$th roots of
unity. The correspondence $$g^\vee\longmapsto \left(g^\vee
(e_1),\ldots, g^\vee(e_1)\right)$$ yields an isomorphism
$$\varphi: G^\vee\stackrel{\cong}{\longrightarrow}\mu_{\bar n} \,.$$
This isomorphism relates the symbol of a convolution operator
$\D_a$ and the Fourier transform $\Hat a$ of its kernel.

\blem\label{syb1} For any $a\in\cF(G,\bar K)$ we have $$\Hat a=
(\Symb_{a,\cV} \vert \mu_{\bar n})\circ\varphi\,.$$ Consequently
\be\label{muva} \charpol_{a,\bar n,\cV}:=\charpol_{a,\Lambda_{\bar
n,\cV}} =\prod_{\xi\in\mu_{\bar n}} (x-\Symb_{a,\cV}(\xi))\,.\ee
\elem

\bproof For any $g\in G, g^\vee\in G^\vee$, letting $\xi_i=g^\vee
(e_i)$, $i=1,\ldots, s$, by (\ref{decoshi}) and (\ref{symbol}) we
obtain:
$$\Hat a(g)\cdot g^\vee (g)=\D_a (g^\vee) (g)=\left(\sum_{v\in G}
a(v)\tau_{-v}\right)(g^\vee) (g)= \sum_{v\in G}
a(v)\tau_{-v}(g^\vee) (g)$$
$$=
\sum_{v=\sum_{j=1}^s \alpha_je_j\in G} a(v)g^\vee(g-v)=
\Symb_{a,\cV} (\xi_1,\ldots,\xi_s)\cdot g^\vee (g)=\Symb_{a,\cV}
(\xi)\,,$$ where $\xi=(\xi_1,\ldots,\xi_s)\in \mu_{\bar n}$.
Indeed
$$g^\vee(g-v)=g^\vee(g)g^\vee(-v)=g^\vee(g)g^\vee
\left(-\sum_{j=1}^s \alpha_je_j\right) =g^\vee(g)\prod_{j=1}^s
\xi_i^{-\alpha_i}\,.$$ The equality (\ref{muva}) follows now from
Proposition \ref{futra4}. \eproof

\subsection{Chebyshev-Dickson systems and
iterated resultants} \bdefi\label{syb2} We consider a Laurent
polynomial $\Omega=P/y^\alpha$, where $P\in\bar K[y_1,\ldots,y_s]$
is a polynomial coprime with
$y^\alpha=y_1^{\alpha_1}\cdot\ldots\cdot y_s^{\alpha_s}$. Given a
multi-index $\bar n=(n_1,\ldots,n_s)\in\N^s$, we define
recursively the iterated resultant $\res_{\bar n}
(\Omega):=Q_s\in\bar K[x]$  via
$$Q_0(x,y_1,\ldots,y_s)=y^\alpha x-P(y_1,\ldots,y_s)$$ and
$$
Q_i(x,y_{i+1},\ldots,y_s)=\res_{y_i} \left(Q_{i-1}(x,y_{i},
\ldots,y_s),y_i^{n_i}-1\right)\in\bar
K[x,y_{i+1},\ldots,y_s],\quad i=1,\ldots,s\,.$$ In detail
$$\res_{\bar n} (\Omega)=\res_{y_s}\left(\ldots
\res_{y_1}\left(y_1^{\alpha_1}\ldots y_s^{\alpha_s}
x-P(y_1,\ldots,y_s), y_1^{n_1}-1\right), \ldots,
y_s^{n_s}-1\right)\,.$$ Clearly
$\lambda=\Omega(\xi)=P(\xi)/\xi^\alpha$ for some $\xi\in\mu_{\bar
n}$ if and only if $\res_{\bar n} (\Omega) (\lambda)=0$.\edefi

Given a lattice $\Lambda$ of rank $s$, a basis $\cV$ of $\Lambda$
and a multi-index $\bar n\in\NC^s$, we let as before
$\Lambda_{\bar n, \cV}$ denote the product sublattice $\bigoplus
n_i\Z v_i$ of $\Lambda$. In the $p$-free case we deduce from Lemma
\ref{syb1} the following expression for the Chebyshev-Dickson
polynomial $\charpol_{a,\bar n,\cV}$ in (\ref{muva}) as the
multivariate iterated resultant of the symbol $\Symb_{a,\cV}$ in
(\ref{symbol}). We leave the details to the reader.

\bprop\label{syb3} In the notation as above, the characteristic
polynomial of the restriction $\D_a\vert \Lambda_{\bar n,\cV}$,
where $a\in\cF(\Lambda,\bar K)$ and $\bar n\in\NC^s$, can be
expressed as follows:
$$\charpol_{a,\bar n,\cV}=
{\Res}_{\bar n} (\Symb_{a,\cV}) \,.$$ \eprop

Cf. \cite[3.3]{HMP}  for an alternative expression of the
characteristic polynomials of $\sigma^+$-automata on
multi-dimensional grids in terms of iterated resultants.

\bexa\label{syb13} Using the above proposition we derive the
following expression for the classical Chebyshev-Dickson
polynomials $T_n$ of the first kind:
$$T_n(x)=\res_y(xy+y^2+1,y^n+1)\,.$$ \eexa

\brem\label{syb14}  Despite the fact that any Chebyshev-Dickson
system satisfies the divisibility property, its individual members
can be arbitrary polynomials. Let us show for instance  that given
a degree $d>0$ polynomial $P\in \bar K[x]$, where $K=\GF(q)$,
there exists a function $a\in \cF^0(\Z,\bar K)$ such that
$P=\charpol_{a, d, e_1}$. Indeed enumerating arbitrarily the roots
$z_1,\ldots,z_d\in\bar K$ of $P$ we consider a function $\Hat a_*$
on $G^\vee$ with $\Hat a_* (i)=z_{i+1}$, $i=0,\ldots,d-1$, where
$G=\Z/d\Z$. Letting $a_*=F^{-1}(\Hat a_*)\in\cF(G,\bar K)$ we push
$a_*$ backward  via $\Z\twoheadrightarrow G$ to a function $a\in
\cF^0(\Z,\bar K)$ supported on the interval $[0,\ldots,d-1]$. Then
$a$ is as required. \erem

\section{Counting points on symbolic variety}
\subsection{Harmonic characters as points on symbolic variety}
We let as before $K=\GF(p^r)$. Given $\bar
a=(a_1,\ldots,a_t)\in(\cF^0(\Lambda,\bar K))^t$, we establish in
Proposition \ref{equa} below a natural bijection between the set
of points on the symbolic variety $\Sigma_{\bar a,\cV}$ in
(\ref{symbvar}) and the set $\Char_{\bar a-{\rm
harm}}(\Lambda,\bar K^\times)$ of all $\bar a$-harmonic characters
of $\Lambda$.

Since $\bar K^\times$ is a torsion group, given a basis
$\cV=(v_1,\ldots,v_s)$ of $\Lambda$, any character $g^\vee\in
\Char(\Lambda,\bar K^\times)$ is $(\bar n,\cV)$-periodic for $\bar
n=(n_1,\ldots,n_s)$, where $n_i=\ord\, (g^\vee (v_i))\in\NC$,
$i=1,\ldots,s$. Letting $G=G_{\bar n,\cV}=\Lambda/\Lambda_{\bar
n,\cV}$ we have $g^\vee=h^\vee\circ\pi$ for a character $h^\vee\in
G^\vee$, where $\pi:\Lambda\twoheadrightarrow G$. By virtue of the
Pushforward Lemma \ref{strf}, $g^\vee$ is $\bar a$-harmonic if and
only if $h^\vee$ is $\bar a_*$-harmonic. Consequently
$$\Char_{\bar a-{\rm harm}}(\Lambda,\bar K^\times)
=\bigcup_{\bar n\in \NC^s} (G_{\bar n,\cV})^\vee_{\bar a_*-{\rm
harm}}\,.
$$
For any $\bar a=(a_1,\ldots,a_t)\in (\cF^0(\Lambda,\bar K))^t$ and
$\bar n=(n_1,\ldots,n_s)\in\NC^s$, we consider the following
over-determined system of algebraic equations, cf. (\ref{symbol}):
\be\label{se} \Symb_{a_j,\cV} (x_1,\ldots,x_s)=0, \quad
x_i^{n_i}=1,\quad i=1,\ldots,s,\,\,\,j=1,\ldots,t\,.\ee We let
$\Sigma_{\bar a,\bar n,\cV}=\Sigma_{\bar a,\cV}\cap\mu_{\bar n}$
denote the set of all solutions of (\ref{se}), or in other words
the set of all points on the symbolic variety $\Sigma_{\bar
a,\cV}$ in (\ref{symbol}) whose multi-torsion orders divide $\bar
n=(n_1,\ldots,n_s)$. The following result yields (b) of Theorem
\ref{manyop} in the Introduction.

\bprop\label{equa} Given a basis $\cV$ of $\Lambda$, the natural
bijection $$\Char (\Lambda,\bar
K^\times)\stackrel{\cong}{\longrightarrow} (\bar K^\times)^s\,$$
restricts to
$$
\Char_{\bar a-{\rm harm}}(\Lambda,\bar K^\times)
\stackrel{\cong}{\longrightarrow} \Sigma_{\bar a,\cV}\,$$ and
further yields the bijections \be\label{eq0} \Sigma_{\bar a,\bar
n,\cV}=\Sigma_{\bar a,\cV}\cap\mu_{\bar n}
\stackrel{\cong}{\longrightarrow} (G_{\bar n,\cV})^\vee_{\bar
a_*-{\rm harm}}\stackrel{\cong}{\longrightarrow}  V(\bar
a_*)\,,\ee where $\bar a_*=\pi_*\bar a$ is the pushforward of
$\bar a$ under the canonical surjection
$\pi:\Lambda\twoheadrightarrow G_{\bar
n,\cV}=\Lambda/\Lambda_{\bar n,\cV}$. Consequently
$$d(\bar a,\Lambda_{\bar n,\cV})=\card\, V(\bar a_*)=
\card \Sigma_{\bar a,\bar n,\cV}\,.$$ \eprop

\bproof For a character $h^\vee\in G_{\bar n,\cV}^\vee$, letting
$g^\vee=h^\vee\circ\pi\in \Char(\Lambda,\bar K^\times)$ and
$\xi_i=g^\vee(v_i)\in\bar K^\times$ we obtain $\xi_i^{n_i}=1$
$\forall i=1,\ldots,s$ (indeed $n_iv_i\in \Lambda_{\bar
n,\cV}\,\,\forall i$). Moreover $h^\vee\in (G_{\bar
n,\cV})^\vee_{\bar a_*-{\rm harm}}$ if and only if $\forall
j=1,\ldots,t$,
$$h^\vee*(a_j)_*=0\quad\Longleftrightarrow \quad g^\vee*a_j=0
\quad\Longleftrightarrow \quad g^\vee*\left(\sum_{v\in L} a_j(v)
\delta_v\right) =0$$$$ \quad\Longleftrightarrow \quad
\sum_{v=\sum_{i=1}^s \alpha_i v_i\in L} a_j(v)
(g^\vee)^{-1}(v)=0\quad\Longleftrightarrow \quad \Symb_{a_j,\cV}
(\xi_1,\ldots,\xi_s)=0\,,$$ and so $\xi=(\xi_1,\ldots,\xi_s)\in
\bigcap_{j=1}^t \Sigma_{a_j,\cV}=\Sigma_{\bar a,\cV}$. Vice versa,
given a solution $\xi=(\xi_1,\ldots,\xi_s)\in (\bar K^\times)^s$
of (\ref{se}), letting $g^\vee(v_i)=\xi_i$ defines an $(\bar
n,\cV)$-periodic character $g^\vee\in \Char(\Lambda,\bar
K^\times)$. By the same argument as above, $g^\vee$ and the
pushforward character $h^\vee=\pi_*(g^\vee)\in (G_{\bar
n,\cV})^\vee$ are $\bar a$- and $\bar a_*$-harmonic, respectively.
The correspondence
$\xi=(\xi_1,\ldots,\xi_s)\quad\longleftrightarrow\quad h^\vee$
yields the first bijection in (\ref{eq0}). As for the second one,
see \ref{futra4}(b).\eproof

\subsection{Criteria of harmonicity}
We let as before $K=\GF(p)$. Given a basis $\cV$ of $\Lambda$, a
sequence $\bar a\in (\cF^0(\Lambda,\bar K))^t$ and a multi-index
$\bar n\in\NC^s$, we let $G=\Lambda/\Lambda_{\bar n,\cV}$ and
$\bar a_*=\pi_*\bar a$, where $\pi:\Lambda\twoheadrightarrow G$.
We fix a minimal $q_0=q(\bar a_*)=p^{r_0}$ ($r_0>0$) such that
$\widehat{a}_j(G^\vee)\subseteq \GF(q_0)$ $\forall j=1,\ldots,t$.
The preceding results lead to the following criteria.

\bthm\label{MFT} \begin{enumerate}\item[(a)] With the notation as
above, the following conditions are equivalent. \item[(i)] There
exists a nonzero $(\bar n,\cV)$-periodic\footnote{I.e., stable
under the shifts by elements of $\Lambda_{\bar n,\cV}$.} $\bar
a$-harmonic function on $\Lambda$. \item[(ii)] $V(\bar
a_*):=\bigcap_{j=1}^t V(a_{j*})\neq\emptyset$. \item[(iii)] The
system (\ref{se}) has a solution $\xi=(\xi_1,\ldots,\xi_s) \in
\Sigma_{\bar a,\bar n,\cV}=\Sigma_{\bar a,\cV}\cap \mu_{\bar n}$.
\item[(b)] Furthermore $\ker (\D_{\bar a_*})\subseteq
\left(\cF(G,\bar K),*\right)$ coincides with the principal
convolution ideal generated by the function $\prod_{j=1}^t
\left(\delta_e-\D_{(a_{j)_*}}^{q_0-1} (\delta_e)\right)$, and
$$\prod_{j=1}^t \left(1_G-\D_{(a_{j})_*}^{q_0-1}\right) :
\cF (G,\bar K)\twoheadrightarrow \ker (\D_{\bar a_*})$$ is an
orthogonal projection. \item[(c)] For $t=1$ and $a_1=a$, (i)-(iii)
are equivalent to every one of the following conditions:
\item[(iv)] $(\D_{a_*})^{q_0-1}(\delta_e)\neq \delta_e$ or,
equivalently, $(\D_{a_*})^{q_0-1}\neq 1_G $. \item[(v)] The
sequence $\left(\D^k_{a_*} (\delta_e)\right)_{k\ge
0}\subseteq\cF(G,\bar K)$ is not periodic.
\end{enumerate}
\ethm

\bproof The equivalences (i)$\Longleftrightarrow
$(ii)$\Longleftrightarrow $(iii) follow immediately from
\ref{futra4} and \ref{equa}. Since the function $\widehat{(a_j)_*}
\in {\cF} (G^\vee, \bar K)$ takes values in the field $\GF(q_0)$
we have $
\delta_{V(\widehat{(a_j)_*})}=1-\widehat{(a_j)_*}^{q_0-1}$. For
every $j=1,\ldots,t$ the Fourier transform sends
$1_G-\D_{(a_j)_*}^{q_0-1}$ to the operator of multiplication by
$\delta_{V((a_j)_*)}$ in $\cF(G^\vee,\bar K)$, which coincides
with the orthogonal projection onto the corresponding principal
ideal. This yields (b) and (c). \eproof

\brems\label{algo} 1. For $t=1$ and $a=a_1$ we have
$\widehat{a_*}^{q_0}=\widehat{a_*}$ and so $\D_{a_*}^{q_0+1}
(\delta_e)=\D_{a_*} (\delta_e)=a_*$. Consequently the truncated
sequence $\left(\D^k_{a_*} (\delta_e)\right)_{k\ge 1}$ (which
starts with $a_*$) is periodic with period $l$ dividing $q_0-1$.
Whereas the sequence in (iv) (which starts with $\delta_e$) is
periodic if and only if  $\Lambda$ does not admit a nonzero
$a$-harmonic $(\bar n,\cV)$-periodic function. In the latter case
$\D_{a_*}$ is invertible of finite order in the group $\Aut
(\Lambda/\Lambda_{\bar n,\cV},\bar K)$.

2.  For $K=\GF(2)$ and $G=\Z/n\Z$, according to \cite[1.1.7]{J} or
\cite{MOW}, we have
$$K[G]^{\times}=\left(K[x]/(x^n-1)\right)^{\times} \cong \Z/\nu\Z\,,$$
where
$$\nu=\nu (n)=2^n\prod_{d\vert n}
\left(1-\frac{1}{2^{f(d)}}\right)^{g(d)}\,$$ with
$f(n)=\ord_n(2)=\min\{j\,:\,2^j\equiv 1\mod n\}$ and $g(n)=\varphi
(n)/f(n)$,. Here $\varphi$ stands for the Euler totient function.
We recall that $G$ admits a nonzero $a^+$-harmonic function if and
only if $n\equiv 0\mod 3$. Otherwise the minimal period $l$ as in
(1) above coincides with the order of $\tilde a^+$ in the cyclic
group $\Z/\nu\Z$, so $l\mid \nu$.

3. For $K=\GF(2)$, $t=1$ and $a=a^+$, $\D_{a_*}^{q_0-1} : \cF
(G,\bar K)\twoheadrightarrow  (\ker (\D_{a_*}))^\bot$ is the
orthogonal projection onto the space $(\ker (\D_{a_*}))^\bot$ of
all winning patterns of the 'Lights Out` game on the toric grid
$G=\Z_{\bar n}$; see \cite[\S 2.8]{Za1} or \S\ref{intro1} in the
Introduction.\erems

\section{Convolution equations over finite fields}
We fix a Galois field $K=\GF(q)$ with $q=p^r$ and a finite abelian
group $G$ of order coprime with $p$. Let  $\D_{\bar a}=(\D_{
a_1},\ldots,\D_{a_t})$ be a system of convolution operators with
kernel $\bar a\in(\cF(G,K))^t$. Clearly the dimension of the space
of solutions $\ker (\D_{\bar a})$ is the same in $\cF(G,K)$ and in
$\cF(G,\bar K)$. We show in Theorem \ref{many} below that,
moreover, the former subspace can be recovered  by taking traces
of $\bar a$-harmonic characters (with values in $\bar K$).

We let $\phi_q: \xi\longmapsto \xi^q$ denote the Frobenius
automorphism of $\bar K=\overline {\GF(q)}$.  By the same letter
we denote the induced action $\phi_q:f\longmapsto f^{\phi_q}=f^q$
on the function spaces $\cF(G,\bar K)$ and $\cF(G^\vee,\bar K)$,
respectively.

The rings of invariants
$$[\cF(G,\bar K)]^{\phi_q}=\cF(G,K)\quad\mbox{and}\quad [\cF(G^\vee,\bar
K)]^{\phi_q}=\cF(G^\vee,K)$$ do not correspond to each other under
the Fourier transform. The restriction $D_q=\phi_q\vert G^\vee$ to
the image of $G^\vee\hookrightarrow\cF(G,\bar K)$ is just the
multiplication by $q$ in the abelian group $G^\vee$. We keep again
the same symbol $D_q$ for the induced automorphism of the function
space $\cF(G^\vee,\bar K)$. The latter one being different from
$\phi_q$, we let $\alpha_q$ denote the automorphism $\phi_q\circ
(D_q)^{-1}$ of $\cF(G^\vee,\bar K)$ measuring this difference. In
the next simple lemma (cf. \cite{J}) we show that $\alpha_q$ is
the Fourier dual of the Frobenius automorphism acting on
$\cF(G,\bar K)$.

\blem\label{gf1} The automorphism $\alpha_q\in\Aut
(\cF(G^\vee,\bar K))$ is the Fourier dual of $\phi_q\in
\Aut(\cF(G,\bar K))$. Hence the Fourier image
$F\left(\cF(G,K)\right)$ coincides with the subalgebra
$(\cF(G^\vee,\bar K))^{\alpha_q} \subseteq\cF(G^\vee,\bar K)$ of
$\alpha_q$-invariants. \elem

\bproof For any $f\in \cF(G,\bar K)$ we have
$$\left(\Hat{f}\right)^{\phi_q}=\widehat{f^{\phi_q}}\circ D_q\,.$$
Indeed, for any $g^\vee\in G^\vee$,
$$\left(\Hat f (g^\vee)\right)^q= \left(\sum_{v\in G} f(v)g^\vee
(v)\right)^q =\sum_{v\in G}f^{\phi_q}(v)
(g^\vee)^{\phi_q}(v)=\widehat{f^{\phi_q}}((g^\vee)^{\phi_q})\,.$$
 Therefore
$$f\in \cF(G,K)\,\Longleftrightarrow\,
f=f^{\phi_q}\,\Longleftrightarrow\,
\Hat{f}=\widehat{f^{\phi_q}}\quad\Longleftrightarrow\quad \Hat{f}
\circ D_q= (\Hat{f})^{\phi_q}\,\Longleftrightarrow\, \Hat{f}
=\alpha(\Hat{f})\,,$$ as stated. \eproof

Now one can easily deduce the following fact.

\bcor\label{gf2} For any $\bar a\in (\cF(G,K))^t$,  the locus
$V(\bar a)\subset G^\vee$ of $\bar a$-harmonic characters is
$D_q$-stable .\ecor

This leads to a direct sum decomposition of the space of
solutions, see \ref{inve}(b) below. For a function $f\in\cF(G,\bar
K)$ we let $GF(q(f))$, where $q(f)=q^{r(f)}$, denote the minimal
subfield of $\bar K$ generated by $K$ and by the image $f(G)$. The
trace of $f$ is
$$\Tr(f)=\Tr_{GF(q(f)):GF(q)}(f)=f+f^q+\ldots+f^{q^{r(f)-1}}\in\cF(G,K)\,.$$

\bprop\label{inve} For any $\bar a\in (\cF(G,K))^t$ the following
hold.
\begin{enumerate}
\item[(a)] There is a bijection between the set of traces $\Tr
(g^\vee)\in\cF(G,K)$ of all $\bar a$-harmonic characters
$g^\vee\in V(\bar a)$ and the orbit space $V(\bar a)/\langle
D_q\rangle$ of the cyclic group $\langle D_q\rangle$ acting on
$V(\bar a)$. \item[(b)] Given a set of representatives
$g_1^\vee,\ldots,g_m^\vee$ of the $\langle D_q\rangle$-orbits on
$V(\bar a)$, there is a decomposition into orthogonal direct sum
of convolution ideals
$$\ker (\D_{\bar a})=\bigoplus_{i=1}^m \left(\Tr (g_i^\vee)
\right)\subseteq \cF(G,K)\,.$$ \item[(c)] For any $ g^\vee\in
G^\vee$ one has
$$g^\vee=\frac{1}{\ord(G)}\sum_{g\in G}
g^\vee (g^{-1}) h_g,\quad\mbox{ where}\quad h_g(x)=\Tr (g^\vee
(gx))=\tau_g\left(\Tr (g^\vee (x))\right)\,.$$
\end{enumerate}
\eprop

\bproof By virtue of \ref{gf2} for any character $ g^\vee\in
V(\bar a)$ we have
$$h=\Tr (g^\vee)=g^\vee+(g^\vee)^q+\ldots
+(g^\vee)^{q^{r(g^\vee)-1}}\in \ker (\D_{\bar a})\cap \cF(G,K)
\,.$$ Letting $$O(g^\vee)=\{g^\vee,\,(g^\vee)^q,\ldots,
(g^\vee)^{q^{r(g^\vee)-1}}\}$$ be the orbit of $g^\vee$ under the
action of the cyclic group $\langle D_q\rangle$ on $V(\bar a)$,
one can easily deduce that $\card\,(O(g^\vee))=r(g^\vee)$ and, by
 (\ref{FuTr}),
$$\widehat h= \ord(G)\sum_{i=0}^{r(g^\vee)-1}
\delta_{(g^\vee)^{-q^i}}=\ord(G)\delta_{O((g^\vee)^{-1})}\,.$$ Now
$h\quad\longleftrightarrow\quad  O((g^\vee)^{-1})$ is the
correspondence required in (a). In turn (a) implies (b). Whereas
(c) follows by using the orthogonality relations for characters.
Indeed by virtue of (\ref{FuTr}), for any $x\in G$  one has
$$\sum_{g\in G} g^\vee
(g^{-1}) h_g(x)=\sum_{g\in G} g^\vee (g^{-1})
\sum_{i=0}^{r(g^\vee)-1} (g^\vee)^{q^i} (gx)$$ $$=
\sum_{i=0}^{r(g^\vee)-1} \left(\sum_{g\in G} (g^\vee)^{-q^i}
(g^{-1}) g^\vee (g^{-1})\right)(g^\vee)^{q^i}
(x)=\sum_{i=0}^{r(g^\vee)-1} \widehat {(g^\vee)^{-q^i}}
(g^\vee)(g^\vee)^{q^i} (x)$$
$$=\ord(G)\sum_{i=0}^{r(g^\vee)-1}
\delta_{(g^\vee)^{q^i}} (g^\vee)(g^\vee)^{q^i} (x)=\ord(G)g^\vee
(x)\,.$$ \eproof

 The following result is straightforward from
  \ref{equa} and \ref{inve}(c).

  \bthm\label{many} \begin{enumerate}\item[(a)]
  For a finite abelian group $G$
  of order coprime to $p$ and for any
  $\bar a\in (\cF(G,K))^t$, where $K=\GF(p^r)$, the kernel
  $\ker \left(\D_{\bar a}\right)$ is spanned over $K$
  by the shifts of traces of $\bar a$-harmonic
  characters $g^\vee\in G^\vee_{\bar a-{\rm harm}}$.
\item[(b)] Similarly, for any sublattice $\Lambda'\subseteq
\Lambda$ of finite index coprime to $p$ and for any $\bar a\in
(\cF(\Lambda,K))^t$, the kernel $\ker \left(\D_{\bar a}\vert
\cF_{\Lambda'}(\Lambda,K)\right)$ is spanned over $K$
  by the shifts of traces of $\bar a$-harmonic
  characters $g^\vee\in
  \Char_{\bar a-{\rm harm}}(\Lambda,\bar K^\times)$
  with $\Lambda'\subseteq\ker (g^\vee)$.\end{enumerate}
  \ethm

\section{Characteristic sublattices and translation invariant subspaces}
\subsection{Lattice characters with values in $\bar K^\times$}
Given a lattice $\Lambda$, a character of $\Lambda$ with values in
$\bar K^\times$ is just a homomorphism $g^\vee : \Lambda\to \bar
K^\times$. It is called $\bar a$-harmonic, where $\bar
a=(a_1,\ldots,a_t)\in (\cF^0(\Lambda,\bar K))^t$, if $\D_{a_j}
(g^\vee)=0$ $\forall j=1,\dots,t$. We let as before
$\Char(\Lambda,\bar K^\times)$ denote the set of all characters on
$\Lambda$ with values in $\bar K^\times$ and $\Char_{\bar a-{\rm
harm}}(\Lambda,\bar K^\times)$ the set of all $\bar a$-harmonic
characters.

\bexa\label{double} For $K=\GF(2)$, $\Lambda=\Z^2$, $t=1$,
$a_1=a^+$ and for a primitive cubic root of unity
$\omega\in\mu_3$,
$$\theta= \begin{pmatrix}
  \cdots & \cdots & \cdots & \cdots & \cdots & \cdots & \cdots\\
  \cdots & 1 &  \omega& \omega^2 & 1 & \omega & \cdots \\
  \cdots & 1 & \omega & \omega^2 & 1 & \omega & \cdots \\
  \cdots & 1 & \omega & \omega^2 & 1 & \omega & \cdots \\
  \cdots & \cdots & \cdots & \cdots & \cdots & \cdots & \cdots
\end{pmatrix}\,$$ is an $a^+$-harmonic character with values in
$\mu_3\subseteq\bar K^\times$. \eexa

\brems\label{charnew} 1. Clearly $\Lambda/\ker \,(\theta)\cong
\Z/m\Z$ for a character $\theta \in \Char (\Lambda,\bar K^\times)$
if and only if $\ord \,(\theta)=m$. Vice versa given a sublattice
$\Lambda'\subseteq \Lambda$  with $\Lambda/\Lambda'\cong \Z/m\Z$,
where $m\in\NC$, we have $\Lambda'=\ker\, (\theta)$ for the
character $\theta$ on $\Lambda$ defined via
$$\theta:\Lambda\twoheadrightarrow
\Lambda/\Lambda'\stackrel{\cong}{\longrightarrow}\Z/m\Z
\stackrel{\cong}{\longrightarrow} \mu_m\hookrightarrow\bar
K^\times\,.$$

2. We note that $\Lambda(\theta')=\varepsilon (\Lambda(\theta))$
for two characters
  $\theta,\,\theta'\in \Char (\Z^s,\bar K^\times)$ and for
   some
  $\varepsilon\in\SL(s,\Z)$ if and only if $\ord\,(\theta')=\ord\,(\theta)$.
  In the latter case
  $\theta'=\delta\circ \theta\circ\varepsilon$, where
  $\delta\in\Aut (\Z/m\Z)\cong (\Z/m\Z)^\times$
  with $m=\ord\,(\theta)$.
\erems

Fixing a basis $\cV=(v_1,\ldots,v_s)$ of $\Lambda$, for a lattice
vector $v=\sum_{i=1}^s \alpha_i(v)v_i\in \Lambda$ we let $\vec
v=(\alpha_1,\ldots,\alpha_s)\in\Z^s$, where
$\alpha_i=\alpha_i(v)$. Below $\langle\cdot,\cdot\rangle$ stands
for the standard bilinear form on $\Z^s$. We observe the
following.

\bprop\label{hacr}
\begin{enumerate}\item[(a)]
Given a primitive $m$th root of unity $\zeta\in\mu_m$ ($m\in\NC$)
and a primitive lattice vector $v_0\in \Lambda$, the formula
\be\label{hacr11} \theta (v)=\zeta^{\langle \vec v, \vec{v_0}
\rangle}\ee defines a character $ \theta\in\Char (\Lambda,\bar
K^\times)$ of order $m$. \item[(b)] Vice versa any character $
\theta\in\Char (\Lambda,\bar K^\times)$ can be written as in
(\ref{hacr11}) for a suitable primitive $m$th root of unity
$\zeta\in\mu_m$ of order $m=\ord\,(\theta)\in\NC$ and a suitable
primitive lattice vector $v_0=v_\theta=\sum_{i=1}^s \alpha_i
(v_\theta)v_i\in \Lambda$ with $\alpha_i(v_\theta)\in
\{0,\ldots,m-1\}$. \item[(c)] Consequently the period lattice of
$\theta$:
$$\Lambda(\theta)=\ker (\theta)=\{v\in \Lambda\,:\, \langle \vec v,
\vec{v_\theta} \rangle\equiv 0\mod m\}\,$$ is an  index $m$
sublattice of $\Lambda$. \item[(d)] The character $\theta$ as in
(\ref{hacr11}) is $a$-harmonic, where $a\in\cF^0(\Lambda,\bar K)$,
if and only if
  \be\label{hacr1} (\theta * a) (0)=\sigma_{a,\cV} (\xi)=0\quad\mbox{that is}
  \quad \xi\in\Sigma_{a, \cV}
  \,,\ee where
   $\xi=(\zeta^{\alpha_1(v_\theta)},\ldots,\zeta^{\alpha_s(v_\theta)})$ and
   $\sigma_{a, \cV}$ is the symbol of $\D_a$ w.r.t. the basis $\cV$.
\end{enumerate}\eprop

\bproof The proof of (a) is straightforward. To show the converse
we let  $\xi_i=\theta ( e_i)\in \bar K^\times$ and
$n_i=\ord\,(\xi_i)$, $i=1,\ldots,s$. We let also $m=m(\theta)=\lcm
(n_1,\ldots,n_s)$ be the exponent of the group $\mu_{\bar n}$. For
a primitive $m$th root of unity $\tilde{\zeta}\in \mu_m$ we write
$\xi_i=\tilde{\zeta}^{b_i}$, where $n_i=m/\gcd (b_i,m)$. Letting
further $d=\gcd (b_1,\ldots,b_s)$, $b_i=d\alpha_i$ and $d_i=\gcd
(b_i,m)$ we obtain $\gcd (d_1,\ldots,d_s)=1$ and so $\gcd
(d,m)=1$. Hence $\zeta=\tilde{\zeta}^d$ is again a primitive $m$th
root of unity and $\xi_i=\zeta^{\alpha_i}$, $i=1,\ldots,s$.
Therefore $\theta:v\longmapsto \zeta^{\langle \vec v, \vec
{v_\theta}\rangle}$ for a primitive vector $
v_\theta=\sum_{i=1}^s\alpha_i(v_\theta)v_i\in \Lambda$, where
$\alpha_i\in \{0,\ldots,m-1\}$, $i=1,\ldots,s$.

Now (\ref{hacr1}) follows from the equalities
$$\theta*a (0)=\sum_{u\in \Lambda} \theta (u)a(-u)=
\sum_{u\in \Lambda} a(u)\zeta^{-\langle \vec{u}, \vec{v_\theta}
\rangle}$$
$$=\sum_{u\in L} a(u)\zeta^{- \alpha_1(u)
\alpha_1(v_\theta)}\cdot\ldots\cdot\zeta^{-\alpha_s(u)
\alpha_s(v_\theta)}=\sigma_{a,\cV} (\xi)\,.$$\eproof

\bexa\label{siex} A character $\theta\in\Char (\Z^s,\bar
K^\times)$,
  $v\longmapsto \zeta^{\langle v, v_0\rangle}$ as in (\ref{hacr11}),
  where $K=\GF(2)$,
  is
  $a^+$-harmonic if and only if
  \be\label{hacr0} (\theta * a^+) (0)=1+\sum_{i=1}^s
  \left(\zeta^{\alpha_i(v_0)}+\zeta^{-\alpha_i(v_0)}\right)=0\,.\ee\eexa

  \brem\label{siex0} A primitive lattice vector $v_0\in \Lambda$
  will be called $a$-exceptional if the
  symbol $\sigma_{a_*}\in\bar K[x,x^{-1}]$
  is a Laurent monomial, where
  $a_*=\pi_*a\in\cF(\Z,\bar K)$ for the surjection
  $\pi: \Lambda\twoheadrightarrow \Z$, $v\longmapsto
  \langle\vec v,\vec v_0\rangle$.
  Clearly
  there exist many non $a$-exceptional vectors $v_0\in \Lambda$
  as soon as $\card \,(\supp (a))\ge 2$.
  For such a vector $v_0$ any nonzero root
  $\zeta$ of the Laurent polynomial $\sigma_{a_*}$ gives rise to
  an $a$-harmonic character
  $\theta\in\Char (\Lambda,\bar K^\times)$, $v\longmapsto
  \zeta^{\langle\vec v,\vec v_0\rangle}$.
  \erem

  \subsection{Periodicity of solutions of convolution equations}
  A general convolution equation on a lattice $\Lambda$ of rank $\ge 2$
  does admit aperiodic solutions.
  For instance
  on $\Lambda=\Z^2$ there are aperiodic $a^+$-harmonic functions
  with values in
  $K=\GF(2)$. Indeed consider the strip $S=\Z\times \{0,-1\}\subseteq \Lambda$
  of width 2.
  Any function $f_0: S\to \bar {K}$
   admits a unique
  $a^+$-harmonic extension $f_0\rightsquigarrow f$ to $\Lambda$ given via
  $$f(m,n)=f_0(m,-1)+f_0(m,0)+f_0(m+n,0)+f_0(m-n,0),\quad m>0,$$
  on the upper halfplane
  and symmetrically on the lower one.
  Clearly for a generic $f_0$ this extension $f$
  is aperiodic that is $\Lambda(f)=\{0\}$.
  The space of all such $a^+$-harmonic functions
  is of infinite dimension.

  Furthermore there are bi-periodic
  $a^+$-harmonic functions on $\Lambda$ with arbitrarily
  large pairs of periods,
  see \cite{Za1}.

  However all solutions of convolution equations on rank 1 lattices
  are periodic with a period depending only on the equation.
  Indeed the following holds.

  \blem\label{rk1per} For any $a\in\cF^0(\Z,\bar K)\setminus\{0\}$,
  every $a$-harmonic function $f\in \ker \,(\D_a)$ is $m_a$-periodic for
  a certain $m_a>0$ depending only on $a$.
  Consequently the subspace $$\ker \,(\D_a)
  \subseteq \cF_{\Lambda'}(\Lambda,\bar K),\qquad\mbox{
  where}\quad \Lambda'=m_a\Z\subseteq \Lambda=\Z\,,$$ is of finite dimension.
  \elem

  \bproof Replacing $a$ by $a*\delta_n$ with a suitable $n$ we may suppose that
  $a=\sum_{i=0}^N a(i)\delta_{-i}$ with $N\ge 0$ and $a(0), \,a(N)\neq 0$. For
  any $f\in \ker \,(\D_a)$ we have $$0=f*a (0) =
  a(0)f(0)+a(1)f(1)+\ldots+a(N)f(N)\,.$$ Hence
  $$f(N)=b_0f(0)+\ldots+b_{N-1}f(N-1)\,,$$
  where $b_i=-\frac{a(i)}{a(N)}$ and so
  $b_0\neq 0$. Therefore the linear transformation
  $$\varphi :\A^N_{\bar K}\to
  \A^N_{\bar K}, \qquad
  (f_0,\ldots,f_{N-1})\longmapsto (f_1,\ldots,f_{N})$$
  with $\det (\varphi)=\pm b_0$
  is invertible, hence of finite order, say, $m_a$.
  This shows that $f$ is periodic of period $m_a$.
  \eproof

  For instance any $a^+$-harmonic function on $\Z$ is periodic of period $3$.
  Moreover any
  $a^+$-harmonic function on $\Lambda=\Z^2$ which is periodic on the strip
  $S=\Z\times \{0,-1\}$ is
  also bi-periodic.
  In the same fashion, one can easily prove the following fact.

  \bcor\label{rksper}  Let $\bar a=(a_1,\ldots,a_l)\in
  (\cF^0(\Lambda,\bar K))^l$ and
  suppose that the convolution ideal $(a_1,\ldots,a_l)
  \subseteq \cF^0(\Lambda,\bar K)$
  contains $s=\rk (\Lambda)$ nonzero functions
  $b_1,\ldots,b_s\in \cF^0(\Lambda,\bar K)$ such that
  $\supp (b_j)\subseteq \Z v_j$
  ($j=1,\ldots,s$), where $v_1,\ldots,v_s\in \Lambda$
  are linearly independent.
  Then any function $f\in\ker (\D_{\bar a})=\bigcap_{j=1}^l\ker (\D_{a_j})$
  is
  pluri-periodic and its period lattice $\Lambda(f)$ contains a
  rank $s$ sublattice
  $\Lambda'=\sum_{j=1}^s m_j\Z v_j$, where $m_j>0$, $j=1,\ldots,s$,
  depend only on $\bar a$.
  \ecor

\subsection{Translation invariant subspaces}
 The following simple observation will be useful.
 For any finitely generated abelian group $G$ with a Sylow
 $p$-subgroup $G(p)$ one has
 $$G(p)=\bigcap_{\theta\in\Char (G, \bar K^\times)} \ker (\theta)
 \quad\mbox{and}\quad
 \Char (G, \bar K^\times)=\pi^*\Char (G/G(p), \bar K^\times)\,,$$
 where $\pi: G\twoheadrightarrow  G/G(p)$.

 Let us introduce the following notions.

\bdefi\label{zschar} A subspace $E\subset\cF(\Lambda,\bar K)$
spanned by a finite set of characters $\theta_1,\ldots,\theta_m\in
\Char (\Lambda,\bar K^\times)$ will be called characteristic. We
note that any characteristic subspace is translation invariant.

A sublattice will be called characteristic if it is of the form
$$\Lambda'=\bigcap_{i=1}^m \ker (\theta_i)\,,$$ where
$\theta_1,\ldots,\theta_m\in \Char (\Lambda,\bar K^\times)$.
\edefi

\blem\label{cara} A sublattice $\Lambda'\in\cL$ is characteristic
if and only if $\Lambda'\in\cL^0$.\elem

\bproof If $\Lambda'\subseteq\Lambda$ is  characteristic then
$\Lambda'=\ker (\varphi)$, where
$\varphi=(\theta_1,\ldots,\theta_m): \Lambda\to \bigoplus_{i=1}^m
\mu_{n_i}$ with $n_i=\ord\,(\theta_i)\in\NC$, $i=1,\ldots,m$.
Hence the index of $\Lambda'$ in $\Lambda$ is coprime with $p$
i.e., $\Lambda'\in \cL^0$.

Conversely assuming that $\Lambda'\in\cL^0$, the order of the
group $G=\Lambda/\Lambda'$ is coprime with $p$. Hence $G$ is a
product of cyclic groups $\mu_{m_i}$ of orders $m_i\in\NC$,
$i=1,\ldots,m$. The compositions
$$\Lambda\twoheadrightarrow G\twoheadrightarrow \mu_{m_i},
\quad i=1,\ldots,n\,,$$ yield characters $\theta_i\in \Char
(\Lambda,\bar K^\times)$ with $\Lambda'=\bigcap_{i=1}^n \ker
(\theta_i)$, and so the sublattice $\Lambda'\subseteq\Lambda$ is
characteristic. \eproof

\brem\label{recov} Given a finite group $G$ of order coprime with
$p$ and a function
  $f\in\cF(G,\bar K)$, the subgroup of periods $\Lambda(f)\subseteq
G$ is a characteristic subgroup. It can be recovered by $\supp
(\Hat f)$ as follows:
$$\Lambda(f)=\bigcap_{g^\vee\in\supp (\widehat f)} \ker\,(g^\vee)\,.$$
Indeed for $g\in G$, $f_g=f\,\iff \, \widehat {f}\cdot g=\widehat
{f} \,\iff  \, g\vert\supp (\widehat {f})=1\,\iff  \,
g\in\bigcap_{g^\vee\in\supp (\widehat f)}\ker (g^\vee) \,.$\erem

  Let $E\subseteq \cF(\Lambda,\bar K)$ be a translation invariant
  subspace
  of finite dimension. We notice that
  $E$ consists of pluri-periodic functions. Indeed the restrictions
  to $E$ of any shift has finite order. We let $G(E)$ and $\Lambda(E)$
  denote the image and the kernel,
  respectively, of
  the
  homomorphism $$\rho:\Lambda\to\Aut\,(E)\cong\GL(n,\bar K),\qquad v
  \longmapsto
  \tau_v\,.$$ Clearly $\Lambda(E)$ is the period lattice of $E$
  that is
  $\Lambda(E)=\bigcap_{f\in E} \Lambda(f)$. Since $E$
  is translation invariant, for any $f \in E$ one has
  $$E(f):=\spann \left(\tau_v(f)\right)_{v\in \Lambda}\subseteq
  E\,.$$
  The group $G(E)$ is finite; indeed, this is
  a finitely generated abelian
  torsion group. We let $G_p(E)$ denote the
  Sylow $p$-subgroup of $G(E)$.

  \bprop\label{odd} A translation invariant subspace
  $E\subseteq \cF(\Z^s,\bar K)$ of
  finite dimension  is a characteristic subspace
   if and only if the  sublattice $\Lambda(E)\subseteq \Lambda$
   is\footnote{That is $\Lambda(E)\in\cL^0$,
   see Lemma \ref{cara} above.}.\eprop

  \bproof If $E$ is spanned by characters, say,
  $\theta_1,\ldots,\theta_n$ then
  $$\bigcap_{i=1}^n \ker\,(\theta_i)\subseteq
  \Lambda(f)\quad\forall f \in E\,,$$ hence
  $\Lambda(E)=\bigcap_{i=1}^n \ker\,(\theta_i)$ is
  a characteristic sublattice.

  Conversely suppose that $\ind_\Lambda (\Lambda(E))=\ord\,(G(E))$
  is coprime with $p$.
  We have $E=\pi^* (E')$ for a
  suitable translation invariant subspace
  $E'\subseteq \cF(G(E),\bar K)$.
  By \ref{zschar}, $E'$ is a convolution ideal
  spanned by characters.
  Hence $E$
  is spanned by characters.
  \eproof

  However the period lattice of an $\bar a$-harmonic function
  on $\Lambda$ is not
  necessarily characteristic, as the following example shows.

  \bexa\label{odd1} (cf. \cite[Example 2.33]{Za1})
  Letting $K=\GF(2)$, $t=1$,
  $a_1=a^+$ and
  $G=\Z/3\Z\oplus\Z/6\Z$, the $a^+$-harmonic function on $G$
$$h=\begin{pmatrix}
  1 & 0 & 1 & 1 & 0 & 1 \\
  0 & 0 & 0 & 1 & 1 & 1 \\
  0 & 0 & 0 & 1 & 1 & 1
\end{pmatrix}\,$$ lifts to
  $f=h\circ\pi\in\ker_{a^+} (\Lambda)$ via
  $\pi:\Lambda=\Z^2\twoheadrightarrow G$. Thus $f$ is  $a^+$-harmonic
  and has the period lattice
$\Lambda(f)=\pi^{-1}(\Lambda(h))=3\Z e_1+6\Z e_2\subseteq \Lambda$
of even index.
  By virtue of Proposition \ref{odd}, $f$ cannot be represented as a linear
  combination of characters with values in $\bar K^\times$.
  \eexa

 Any sublattice $\Lambda'\in\cL$ is contained in a unique
 minimal characteristic sublattice
 $\Lambda''\in\cL^0$, where
 $$\Lambda''=\bigcap_{\tilde{\Lambda}\in\cL^0,\,
 \tilde{\Lambda}\supseteq \Lambda'} \tilde{\Lambda}=
 \bigcap_{\theta\in\Char (\Lambda, \bar K^\times),\,
 \ker (\theta)\supseteq \Lambda'}
 \ker (\theta)\,.$$
 These data fit in the
 following commutative diagram:
 \begin{displaymath}
 \begin{diagram}[size=2em,textflow,labelstyle=\scriptscriptstyle]
                    &      & 0      &      &               &           &           &      &  \\
                    &      & \dTo   &      &               &           &           &      &  \\
                  0 & \rTo & \Lambda(E)   & \rTo & \Lambda             & \rTo^\rho & G(E)      & \rTo & 0 \\
                    &      & \dTo   &      & \dTo>{\rm id} &           &  \dTo     &      &   \\
                  0 & \rTo & \Lambda''(E) & \rTo & \Lambda             &\rTo       &G(E)/G_p(E)& \rTo & 0 \\
            &      &        &      &               &           &   \dTo    &      &  \\
                    &      &        &      &               &           &     0     &      &  \\
  \end{diagram}\end{displaymath}
 Any finite dimensional translation invariant subspace
 $E\subseteq\cF(\Lambda,\bar K)$
 contains a unique maximal characteristic subspace
 $E^0\subseteq E$, where
 $$E^0=
 \spann\,\left(\theta \,:\,\theta\in E\cap \Char (\Lambda, \bar
 K^\times)\right)\,.$$
 Actually $E^0$ is the fixed point subspace for the action of
 $G_p(E)$ on $E$ by shifts.
 The next lemma says that in characteristic  $2$ case, $E^0\neq 0$
 whenever $E\neq 0$.

 \blem\label{chper} Suppose that $p=2$, and let
 $E\subseteq \cF(\Lambda,\bar K)$
be a non-trivial translation invariant subspace of finite
dimension. Then there exists a nonzero function $h\in E^0$ such
that the sublattice of periods $\Lambda(h)\subseteq \Lambda$ is
characteristic i.e., of odd index. \elem

\bproof If $2 v\in \Lambda$ is
  a period of a nonzero function $f\in E$
  then so is $v$, maybe, for another such function $h\in E$.
  Indeed since $f=f_{2v}\neq 0$ then either $f=f_{v}=h$ or
  $f\neq f_{v}$, and then $h=f+f_{v}\neq 0\in E$ is $v$-periodic,
  as
  required. In this way we arrive finally at a
  sublattice $\Lambda''=\Lambda(h'')\subseteq \Lambda$
  of odd index which contains $\Lambda(E)$.
This corresponds to passing from the group
$G(E)=\Lambda/\Lambda(E)$ to its quotient group $G(E)/G_2(E)$ of
odd order, where $G_2(E)$ is the Sylow $2$-subgroup of $G(E)$. By
Proposition \ref{odd}, $h''\in E^0$ as required.
  \eproof

For any sublattice $\Lambda'\in\cL$
  the translation invariant
  subspace
  \be\label{maxi1} E_{\Lambda'}=\cF_{\Lambda'}(\Lambda,\bar K)=
  \{f\in \cF(\Lambda,\bar K)\,:\,\Lambda(f)\supseteq \Lambda'\}
  \subseteq\cF(\Lambda,\bar K)\,\ee
  is the maximal such subspace with period lattice $\Lambda'$.
  It is easily seen that
  $\Lambda''=\Lambda(E^0_{\Lambda'})$.
  The regular representation
  $\rho:\Lambda\to \Aut(E_{\Lambda'}),\,\,v\longmapsto \tau_v$,
  factorizes through a representation  $\Lambda\to S_n$ into
  the symmetric group, where $n=\ind_{\Lambda} (\Lambda')=
  \dim_{\bar K} (E_{\Lambda'})$
  and $S_n$
  acts by permutations of the orthonormal basis
  $((\delta_{v+\Lambda'})_{v\in \Lambda}$ of $E_{\Lambda'}$.
  The latter one can
  be identified with the basis of $\delta$-functions in
  $\cF(G,\bar K)\cong E_{\Lambda'}$,
  where $G=\Lambda/\Lambda'$.
The representation $\rho$ is induced via $L\twoheadrightarrow G$
by the regular representation of $G$ on $\cF(G,\bar K)$. The
matrix elements of $\rho$ are the $\delta$-functions $\delta_g$
($g\in G$). Every function $f\in \cF(G,\bar K)$  is a state i.e.,
a linear combination of matrix elements.

\subsection{Examples}
We let below $K=\GF(2)$, $\Lambda=\Z^s$, $t=1$, $a=a^+$, and we
denote $a_*$ again by $a^+$. A sublattice
$\Lambda'\subseteq\Lambda$ will be called $\bar a$-harmonic if
there exists a nonzero $\bar a$-harmonic $\Lambda'$-periodic
function $f:\Lambda\to \bar K$.

\bexa\label{nex} Supposing $s=2$, for a primitive vector
$u_0=(k,l)\in\Lambda=\Z^2$ we let $\Lambda_0=\Z v_0$,
   where $ v_0=(-l,k)\bot u_0$.
   Since $\gcd (k,l)=1$ we have
   $\Lambda/\Lambda_0\cong \Z$. We let $\pi_{0}: \Lambda\twoheadrightarrow
   \Lambda/\Lambda_0\cong \Z$,
   $ u\longmapsto \langle  u, u_0\rangle$.
   The induced operator $\D_{a^+_*}\in\End
   (\cF(\Lambda/\Lambda_0,K))$
   corresponds to the following function on $\Z$:
   $$a^+_*=\delta_0+\delta_k+\delta_{-k}+\delta_l+\delta_{-l}\,.$$
   For a function
   $f\in \cF(\Lambda/\Lambda_0,\bar K)$, one has
   $\tilde f:=f(-lx+ky)\in\ker(\D_{a^+})$ if and only if  $f$
   satisfies the equation
   \be\label{haz} f(z)+f(z-k)+f(z+k)+f(z-l)+f(z+l)=0,
   \qquad\forall z\in
   \Z\,.\ee
   In particular if $ u_0= (0,1)$ then
   $\tilde f\in \ker(\D_{a^+}\vert \Z^2)\quad\iff  \quad
   f\in \ker(\D_{a^+}\mid \Z)$. The only
   $a^+$-harmonic sublattice $\Lambda'\in \cL^0$ that contains the
   vector $e_1$ is $L'=\Z  e_1+3\Z  e_2$ (cf. \ref{siex0} below).

   Further,
   if $ u_0 =\pm  e_1\pm e_2$ then
   $\tilde f\in \ker(\D_{a^+})\quad\iff  \quad f=0$.
   Consequently none of the $a^+$-harmonic sublattices $\Lambda'\in \cL$
   contains a vector of the form
   $\pm  e_1\pm e_2$.

Next we let $\Lambda'=\Z u_0+\Z v_0\subseteq\Z^2=\Lambda$, where
$u_0=(k,l)$ and $ v_0=(k',l')$ are primitive lattice vectors
different from $\pm  e_1,\pm  e_2,\pm  e_1\pm e_2$. Suppose that
$m=\ind_\Lambda (\Lambda')=|\det ( u_0, v_0)|$ is odd. Then
$\Lambda'$ is $a^+$-harmonic if and only if there exists a nonzero
solution $f\in\cF(\Z,\bar K)$ of (\ref{haz}), if and only if there
is a primitive $m$th root of unity $\zeta\in\bar K^\times$
satisfying
$$1+\zeta^k+\zeta^{-k}+\zeta^l+\zeta^{-l}=
0\,.$$ Such a root $\zeta$ determines an $a_*^+$-harmonic
character $\theta\in\Char_{a_*^+-{\rm harm}}
(\Lambda/\Lambda_0,\bar K^\times)$ of order $m$, where $\theta:
x\longmapsto \zeta^x$. It also defines an $a^+$-harmonic character
$\theta\circ\pi_0\in\Char_{a^+-{\rm harm}} (\Lambda,\bar
K^\times)$ and the corresponding sublattice
$\Lambda'=\ker(\theta\circ\pi_0)$.

For a finite abelian group $G=\bigoplus_{i=1}^s \Z/n_i\Z$
 of odd order
we have by \ref{futra4}(b) $$\spec\, (\D_{a^+,G})=\widehat
{a^+}(G^\vee)\subset\bar K\,.$$ Letting $g^\vee\in G^\vee$ be a
character with $g^\vee ( e_j)=\xi_j=\zeta_j^{k_j}$, where
$\zeta_j\in \mu_{n_j}$ is a primitive $n_j$th root of unity and
$0\le k_j\le n_j-1$, we obtain (cf. (\ref{sstar}))
$$\widehat {a^+} (g^\vee)=1+
\sum_{j=1}^s \left(g^\vee ( e_j)+g^\vee ( e_j)^{-1}\right)=1+
\sum_{j=1}^s \left(\zeta_j^{k_j}+\zeta_j^{-k_j}\right)\,.$$
Therefore
$$\charpol (\D_{a^+,G})=
\prod_{(k_1,\ldots,k_s)\in\Z_{\bar n}}\left(x-\left(1+
\sum_{j=1}^s
\left(\zeta_j^{k_j}+\zeta_j^{-k_j}\right)\right)\right)\,.$$
\eexa

\bexas\label{four1} (see \cite{Za1}) With $\cV=(e_1,\ldots,e_s)$
being the standard basis in $\Lambda$, the following hold.

\smallskip \noindent $\bullet$
A product sublattice $\Lambda_{\bar n, \cV}$ is $a^+$-harmonic if
and only if the image $\tilde a^+$ of the symbol \be\label{sstar}
\Symb_{a^+,\cV}= 1+\sum_{i=1}^s (x_i+x^{-1}_i)\,\ee is a zero
divisor in $K[G]$.

\smallskip \noindent $\bullet$
If $s=1$ then a sublattice $n\Z\subseteq\Z$ is $a^+$-harmonic
(equivalently, $1+x+x^{-1}$ is a zero divisor in $K[x]/(1+x^n)$)
if and only if $n\equiv 0 \mod 3$.

\smallskip \noindent $\bullet$
Similarly for every $\bar n=(n_1,\ldots,n_s)\in\N^s$ with
$n_1\equiv 0\mod 3$, the group $\Z_{\bar n}$ is $a^+$-harmonic.

\smallskip \noindent $\bullet$
The group $\Z_{5,5}=(\Z/5\Z)^2$ is $a^+$-harmonic, while
$\Z_{7,7}=(\Z/7\Z)^2$ is not.

Many more examples of this kind were computed by L.
Makar-Limanov\footnote{A letter to the author, 2004, 5p.} along
the same lines, and in \cite[Appendix 1]{Za1} by different
methods. Cf. also \ref{5x5} below. \eexas

\bexas\label{sta} ($s=1$) 1.  For $\Lambda'=2\Z\subseteq \Z$ the
representation
$$\rho: \Z\to\Z/2\Z\to S_2\hookrightarrow\GL(2,K),
\qquad 1\longmapsto \beta=\left(\begin{matrix} 0 & 1\\ 1 &
0\end{matrix}\right) \quad\mbox{with}\quad \beta^2=1\,$$  is
equivalent to
$$\rho':1\longmapsto \alpha=\left(\begin{matrix} 1 & 1\\ 0 & 1\end{matrix}
\right) \quad\mbox{with}\quad \alpha^2=1\,.$$ The matrix elements
of $\rho$ provide the 2-periodic function $\delta_\Lambda=(\ldots
0,1,0,1,0,1\ldots)$ on $\Z$ and its shifts, whereas the other
states are constant functions.

2. Similarly for $\Lambda'=3\Z\subseteq \Z$, $\rho$ factorizes
through a faithful representation $\Z/3Z\to S_3$. The matrix
elements give rise to the shifts of the periodic function
$$\delta_{\Lambda'}=(\ldots 0,0,1,0,0,1,0,0,1\ldots)$$ on $\Z$.
The $a^+$-harmonic function $(\ldots 0,1,1,0,1,1,0,1,1\ldots)$ and
its shifts are states.

3. A cyclic group $G=\Z/m\Z$ is $a^+$-harmonic if and only if
$m\equiv 0 \mod 3$, see \ref{four1}.2. Letting $m=3l$, $l\in\N$,
we fix a primitive $m$th root of unity $\zeta\in\mu_{3l}$, a
primitive cubic root of unity $\omega\in\mu_3$, and we let
$g^\vee: n e_1\longmapsto \zeta^n$ be the corresponding character
of $G=\Z/3l\Z$. For $\theta=(g^\vee)^t$ one has
$$\theta\in G^\vee_{a^+-{\rm harm}}\quad\iff  \quad
\zeta^t+\zeta^{-t}=1\quad\iff  \quad \zeta^t= \omega^{\pm
1}\quad\iff \quad t\equiv \pm l\mod 3l.$$ So
$\theta=(g^\vee)^l:\Z/3l\Z\to\F_4^\times $ is an $a^+$-harmonic
character with trace
$$h=\Tr_{\F_4} (\theta) : n e_1\longmapsto
\omega^n+\omega^{2n}=
  \begin{cases}
    0 & \text{if} \,\,n\equiv 0 \mod 3  , \\
    1 & \text{otherwise}\,.
  \end{cases}$$
  Furthermore  $$d_{a^+,G}=2, \quad V(a^+)=G^\vee_{a^+-{\rm
  harm}}=\{\theta,\theta^{-1}\}\quad\mbox{ and}
  \quad \ker(a^+)=\spa (h,h^+)\,,$$
  where
  $h^+(x)=h(1+x)$.

  Similarly for
  $\theta(x)=\omega^x\in \Char_{a^+-{\rm harm}} (\Z,\bar K^\times)$,
  $L(\theta)=3\Z$
  is a maximal proper $a^+$-harmonic sublattice of $\Z$. Moreover
  $h, h^+$ lifted
  to $\Z$ give a basis of $\ker (\D_{a^+}\vert 3\Z)$. \eexas

\bexa\label{5x5}  ($s=2$) As another example, we consider the
group $G=\Z/5\Z\oplus\Z/5\Z$. We fix a primitive $5$th root of
unity $\zeta\in\mu_5$. We have $d_{a^+,G}=8$, see e.g.,
\cite[Appendix 1]{Za1}. The relation
$\zeta+\zeta^2+\zeta^3+\zeta^4=1$ yields the $8$ solutions of
(\ref{se}) with $s=2$, $t=1$, $a_1=a^+$,
$\sigma_{a^+}=x_1+x_1^{-1}+x_2+x_2^{-1}+1$ and $n_1=n_2=5$. These
solutions can be obtained from $(x_1,x_2)=(\zeta, \zeta^3)$ by
suitable transformations.
 The locus of $a^+$-harmonic characters
$$V(a^+)=G^\vee_{a^+-{\rm harm}}=\begin{pmatrix}
  0 & 0 & 0 & 0 & 0 \\
  0 & 0 & 1 & 1 & 0 \\
  0 & 1 & 0 & 0 & 1 \\
  0 & 1 & 0 & 0 & 1 \\
  0 & 0 & 1 & 1 & 0
\end{pmatrix}\,$$
consists of two orbits of the cyclic group $\langle
D_2\rangle\cong \Z/4\Z$ acting on $G^\vee$ via $D_2:
g^\vee\longmapsto (g^\vee)^2$.

The solution $(\zeta, \zeta^3)$ ($(\zeta^3, \zeta)$, respectively)
of (\ref{se}) gives rise to the $a^+$-harmonic character $g^\vee :
m e_1+n e_2\longmapsto \zeta^{m+3n}$ ($\,^t g^\vee : m e_1+n
e_2\longmapsto \zeta^{3m+n}$, respectively), where
$$g^\vee=\begin{pmatrix}
 1 & \zeta^3 & \zeta & \zeta^4 & \zeta^2 \\
  \zeta & \zeta^4 & \zeta^2 & 1 & \zeta^3 \\
  \zeta^2 & 1 & \zeta^3 & \zeta & \zeta^4 \\
 \zeta^3 & \zeta & \zeta^4 & \zeta^2 & 1 \\
  \zeta^4 & \zeta^2 & 1 & \zeta^3 & \zeta
\end{pmatrix}\,\,\,\mbox{resp.}\,\,\,
\,^t g^\vee=\begin{pmatrix}
 1 & \zeta & \zeta^2 & \zeta^3 & \zeta^4 \\
  \zeta^3 & \zeta^4 & 1 & \zeta & \zeta^2 \\
  \zeta & \zeta^2 & \zeta^3 & \zeta^4 & 1 \\
 \zeta^4 & 1 & \zeta & \zeta^2 & \zeta^3 \\
  \zeta^2 & \zeta^3 & \zeta^4 & 1 & \zeta
\end{pmatrix}\,$$
has trace
$$h=\Tr_{\F_{16}}(g^\vee)=\begin{pmatrix}
  0 & 1 & 1 & 1 & 1 \\
  1 & 1 & 1 & 0 & 1 \\
  1 & 0 & 1 & 1 & 1 \\
  1 & 1 & 1 & 1 & 0 \\
  1 & 1 & 0 & 1 & 1
\end{pmatrix}\,\,\mbox{resp.}
\,\,\, \,^t h=
\begin{pmatrix}
  0 & 1 & 1 & 1 & 1 \\
  1 & 1 & 0 & 1 & 1 \\
  1 & 1 & 1 & 1 & 0 \\
  1 & 0 & 1 & 1 & 1 \\
  1 & 1 & 1 & 0 & 1
\end{pmatrix}\,.$$

The doubling of periods as in \cite[2.35]{Za1} produces the
following $a^+$-harmonic function on $G=(\Z/10\Z)^2$:
$$\delta(h)=\begin{pmatrix}
  0 & {\bf 1} & {\bf 1} & 0 & {\bf 1} & 0 & {\bf 1} & 0 & {\bf 1} & {\bf 1} \\
  {\bf 1} & 0 & 0 & 0 & 0 & 0 & {\bf 1} & 0 & 0 & 0 \\
  {\bf 1} & 0 & {\bf 1} & 0 & {\bf 1} & {\bf 1} & 0 & {\bf 1} & {\bf 1} & 0  \\
  0 & 0 & {\bf 1} & 0 & 0 & 0 & {\bf 1} & 0 & 0 & 0  \\
  {\bf 1} & {\bf 1} & 0 & {\bf 1} & {\bf 1} & 0 & {\bf 1} & 0 & {\bf 1} & 0  \\
  0 & 0 & {\bf 1} & 0 & 0 & 0 & 0 & 0 & {\bf 1} & 0 \\
  {\bf 1} & 0 & {\bf 1} & 0 & {\bf 1} & 0 & {\bf 1} & {\bf 1} & 0 & {\bf 1}  \\
  0 & 0 & 0 & 0 & {\bf 1} & 0 & 0 & 0 & {\bf 1} & 0 \\
  {\bf 1} & 0 & {\bf 1} & {\bf 1} & 0 & {\bf 1} & {\bf 1} & 0 & {\bf 1} & 0  \\
  {\bf 1} & 0 & 0 & 0 & {\bf 1} & 0 & 0 & 0 & 0 & 0
\end{pmatrix}\,$$
This matrix contains five crosses $$\begin{pmatrix}
  &  & {\bf 1} & &   \\
  &  & {\bf 1} & & \\
  {\bf 1} & {\bf 1} & 0 & {\bf 1} & {\bf 1}   \\
  &  & {\bf 1} & &   \\
  &  & {\bf 1} & &
\end{pmatrix}\,.$$ Letting $\pi_1:\Lambda=\Z^2
\twoheadrightarrow (\Z/5\Z)^2$ and $\pi_2:\Z^2\twoheadrightarrow
(\Z/10\Z)^2$ it is easily seen that
$\Lambda_1=\Lambda(h\circ\pi_1)=\Z u +\Z v$ and
$\Lambda_2=\Lambda(\delta(h)\circ\pi_2)=2\Z +2\Z v$, where $
u=(1,2)$ and $ v=(-2,1)$. Hence $\ind_{\Lambda} (\Lambda_1)=5$ and
$\ind_{\Lambda} (\Lambda_2)=20$. Actually
$\Lambda_1=\Lambda(\theta)$, where
$\theta=g\circ\pi_1\in\Char_{a^+-{\rm harm}} (\Lambda)$. In
contrast the sublattice $\Lambda_2\subseteq\Z^2$ of even index is
not characteristic (see Definition \ref{zschar}.a), although it is
a period lattice for a nonzero harmonic function on
$\Lambda=\Z^2$. By virtue of Proposition \ref{odd},
$\delta(h)\circ\pi_2$ is not a linear combination of characters
with values in $\bar K^\times$.\eexa

\section{Multi-orders table and Partnership graph}
In this section we let again $K=\GF(q)$, where $q=p^r$ with
$p,r>0$, so that $\bar K^\times$ is a torsion group.

Given a lattice $\Lambda$, a base $\cV$ of $\Lambda$ and a system
$\D_{\bar a}$ of convolution operators on $\Lambda$, where $\bar
a=(a_1,\ldots,a_t)\in\left(\cF^0(\Lambda,\bar K)\right)^t$, we
compose a table $\cD_{\bar a,\cV}= \{d_{\bar a,\bar n,\cV}\}_{\bar
n\in\N^s},$ where $$d_{\bar a,\bar n,\cV}= \dim \ker
\left(\D_{\bar a} \vert \Lambda_{\bar n,\cV}\right)=\dim
\left(\bigcap_{j=1}^t \ker \left(\D_{a_j} \vert \Lambda_{\bar
n,\cV}\right)\right)\,,$$ cf. (\ref{funct}). Thus the entries
$d_{\bar a,\bar n,\cV}$ are nonzero for all $\bar n\in\N^s$ such
that $\Sigma_{\bar a,\bar n,\cV}\neq\emptyset$ (see \ref{MFT}). By
virtue of \ref{equa}, $\forall \bar m=(m_1,\ldots,m_s)\in \NC^s,
\,\forall n=(n_1,\ldots,n_s)\in \NC^s$ \be\label{nonse} d_{\bar
a,\gcd(\bar m,\bar n),\cV}\le \min\, \{d_{\bar a,\bar m,\cV},
\,d_{\bar a,\bar n,\cV}\}\,,\ee where $\gcd(\bar m,\bar
n)=(\gcd(m_1,n_1),\ldots,\gcd(m_s,n_s))$. Indeed due to
(\ref{se}),
$$\Sigma_{\bar a,\gcd(\bar m,\bar n),\cV}=
\Sigma_{\bar a,\bar m,\cV}\cap \Sigma_{\bar a,\bar n,\cV}\,.$$

\brem\label{strict} We note that for $s=1$, the classical
Chebyshev-Dickson polynomials $T_n$ of the first kind satisfy the
identity $\gcd\,(T_m,T_n)=T_{\gcd(m,n)}$. Whereas for $s\ge 2$ the
inequality (\ref{nonse}) is strict in general. For instance for
$K=\GF(2)$, $s=2$, $t=1$ and $a_1=a^+$,
$$d_{a^+,(3,3)}=4<\min\, \{d_{a^+,(9,21)},
\,d_{a^+,(21,9)}\}=16\,.$$ Thus the above identity does not hold
in general for a Chebyshev-Dickson system.\erem

We consider also the table of multi-orders $\cS_{\bar
a,\cV}=\{s_{\bar a,\bar n, \cV}\}_{\bar n\in\NC^s}$, where
$s_{\bar a,\bar n, \cV}=\card \Sigma_{\bar a,\bar n, \cV}$ denotes
the number of points $\xi=(\xi_1,\ldots,\xi_s)$ on the symbolic
hypersurface $\Sigma_{\bar a,\cV}$ with multi-order
\be\label{muor} \bar n=\mord (\xi)=(\ord\, (\xi_1), \ldots,\ord
\,(\xi_s))\in \NC^s\,.\ee These two tables $\cD_{\bar a,\cV}$ and
$\cS_{\bar a,\cV}$ are related via
$$d_{\bar a,\bar n,\cV}=\sum_{\bar d\mid\bar n}
s_{\bar a,\bar d,\cV}\,,$$ where $\bar n\in\NC$ and $\bar d$ runs
over all $s$-tuples $\bar d=(d_1,\ldots,d_s)\in\NC^s$ with
$d_i\mid n_i$ $\forall i=1,\ldots,s$.

Letting $\bar n=(n_1,\ldots,n_s)=(\bar n',n_s)$, for any fixed
$\bar\xi'=(\xi_1,\ldots,\xi_{s-1})\in \mu_{\bar
n'}=\bigoplus_{i=1}^{s-1} \mu_{n_i}$ the system $\sigma_{a_j}
(\bar\xi', x_s)=0$, $j=1,\ldots,t$, has
 a finite set of solutions
$x_s=\eta\in\bar K^\times$. Hence for any $\bar n'$ the line
$(s_{a,\,(\bar n',n_s),\cV})$ of the table $\cS_{\bar a,\cV}$ has
bounded support in $\Z$.

We let $$l(\bar n')=\lcm \left(\ord\, (\eta)\,:\,\exists
\bar\xi'\in \mu_{\bar n'},\, \sigma_{a_j} (\bar\xi',
\eta)=0\quad\forall j=1,\ldots,t\right)\,.$$ Hence for any $\bar
n'$ the line $(d_{a,\,(\bar n',n_s),\cV})_{n_s}$ of the table
$\cD_{\bar a,\cV}$ is periodic with minimal period $l(\bar n')$:
$$d_{a,\,(\bar n',n_s+l(\bar n')),\cV}=d_{a,\,(\bar n',n_s),\cV}\qquad
\forall \bar n'\in \NC^{s-1}\,.$$ The set $(l(\bar n'))_{\bar
n'\in\NC^{s-1}}$ of all such periods is in general unbounded.

For instance for $K=\GF(2)$, $s=2$, $t=1$ and $a_1=a^+$ we have
$$\max_{m\in\NO} \{d_{a^+, (n,m)}\} = d_{a^+, (n,l(n))}=\begin{cases} 2n,
& n\not\equiv 0\mod 3\\2n-2, & \mbox{otherwise}\,.
\end{cases}$$

Following \cite{Za1} we let
$$\cE_{\bar a,\cV}=\{\bar n\in \NC^s\mid
d_{\bar a,\bar n,\cV}\neq 0\}\qquad\mbox{and}\qquad\cE_{\bar
a,\cV}^0 =\{\bar n\in \NC^s\mid s_{\bar a,\bar n,\cV}\neq 0\}\,.$$
By \ref{MFT}, $\cE_{\bar a,\cV}^0\subseteq \cE_{\bar a,\cV}$.
Letting $\bar k\bar n=(k_1n_1,\ldots,k_sn_s)$ we obtain a natural
covering $\pi:\Lambda/\Lambda_{\bar k\bar n,\cV}\to
\Lambda/\Lambda_{\bar {n},\cV}$. Any $\bar a_*$-harmonic function
on the second group lifts to such a function on the first one.
Therefore $\cE_{\bar a,\cV}$ is generated by $\cE_{\bar a,\cV}^0$
as an $\NC^s$-module. So in order to determine $\cE_{\bar a,\cV}$
it is enough to determine $\cE_{\bar a,\cV}^0$.

Assuming that the symbols $\sigma_ {a_j}$, $j=1,\ldots,t$, are
symmetric i.e., stable under the natural action of the symmetric
group $S_s$ on the Laurent polynomial ring $\bar
K[x_1,x_1^{-1},\ldots,x_s,x_s^{-1}]$, it is possible to replace
the multi-orders table $\cS_{\bar a,\cV}$ by the labelled
'partnership hypergraph` $\cP_{\bar a}$. The latter one has the
set of naturals $\NC$ as the set of vertices, and consists of all
$(s-1)$-simplices $\bar n\in\cE_{\bar a,\cV}^0$ labelled with
$s_{\bar a,\bar n, \cV} \neq 0$. For $s=2$, $\cP_{\bar a}$ is just
the infinite labelled graph with the set of vertices $\NC$ and
with the edges $[m,n]\in\cE_{a^+,\cV}^0$ labelled with
$s_{a^+,(m,n), \cV}\neq 0$.

\bexa\label{noue} As was observed by Don Zagier, in the case where
$s=2$, $K=\GF(2)$, $t=1$ and $a_1=a^+$, all connected components
of the partnership graph $\cP_a$ are finite; see \cite[Theorem
3.11]{Za1}. Actually every such component is contained in a level
set $f_0^{-1}(r)$ of the suborder function
$$ f_0(n)={\rm sord}_n (2)
=\min\{j\,:\,2^j\equiv\pm1\mod n\}\,$$ (see e.g. \cite{MOW}).
Indeed for a primitive $n$th root of unity $\zeta\in\mu_n$, by
5.10(a) in \cite{Za1} one has $f_0(n)=\deg\,(\zeta+\zeta^{-1})$.
Let $(\zeta,\eta)\in\Symb_{a^+}$ be a point with bi-order $(\ord\,
(\zeta),\ord\, (\eta))=(m,n)$. Then $(\zeta,\eta)$ satisfies the
symbolic equation
$$\zeta+\zeta^{-1}+\eta+\eta^{-1}=1\,.$$ Hence
$K(\zeta+\zeta^{-1})=K(\eta+\eta^{-1})$ and so
$f_0(m)=f_0(n)$.\eexa

It is plausible \cite[4.1]{Za1} that the connected components of
the partnership graph $\cP_{a^+}$ coincide with the corresponding
level sets of the suborder function $f_0$ except for $r=5$
($f_0^{-1}(5)$ consists of two such components). This conjecture
is based on the computation of the first 13 of these components
done by Don Zagier with PARI, see Appendix 1 in \cite{Za1}.

\smallskip

The following questions arise.

\smallskip

{\bf Problems.}
\begin{enumerate} \item[$\bullet$] Given a lattice $\Lambda$ of rank
$s=2$ and a Galois field $K=\GF(q)$, describe the set of all
functions $a\in\cF^0(\Lambda,K)$ such that all components of the
graph $\cP_a$ are finite.
 \item[$\bullet$]
Is it possible to reconstruct the function $a$ starting from  the
graph $\cP_a$? \item[$\bullet$] Determine the set of all functions
$a\in\cF^0(\Lambda,K)$ such that the irreducible factors of the
Chebyshev-Dickson polynomials $\charpol_{a, \bar n,\cV}$ exhaust
all irreducible polynomials over $K$. This is indeed the case for
the classical Chebyshev-Dickson polynomials $T_n$, see e.g.
\cite[2.8-2.10]{HMP} and the references therein. \item[$\bullet$]
\footnote{This question is proposed by Roland Bacher in the case
of $\sigma^+$-automata.} Does there exist any reasonable
multivariate generating function for the tables $\cD_{\bar a,\cV}$
or $\cS_{\bar a,\cV}$? The natural analogs of the logarithm of the
Weil zeta function is unlikely to play this role.
\end{enumerate}

\end{document}